\pdfoutput=1

\documentclass[sn-mathphys]{sn-jnl}

\jyear{2021}%

\theoremstyle{thmstyleone}%
\newtheorem{theorem}{Theorem}
\theoremstyle{thmstyletwo}%
\newtheorem{remark}{Remark}%

\theoremstyle{thmstylethree}%
\newtheorem{definition}{Definition}%

\usepackage{mathrsfs}
\usepackage{multirow}
\usepackage{enumerate}
\usepackage{subfigure}

\begin{document}

\title[Cooperative Area Coverage and Target Tracking]{Integrated Design of Cooperative Area Coverage and Target Tracking with Multi-UAV System}

\author[1]{\fnm{Mengge} \sur{Zhang}}\email{zhangmg@nudt.edu.cn}

\author[1]{\fnm{Jie} \sur{Li}}\email{lijie09@nudt.edu.cn}
\author*[1]{\fnm{Xiangke} \sur{Wang}}\email{xkwang@nudt.edu.cn}

\affil*[1]{\orgdiv{College of Intelligence Science and Technology}, \orgname{National University of Defense Technology}, \orgaddress{ \state{Changsha}, \country{P.~R.~China}}}






\abstract{This paper systematically studies the cooperative area coverage and target tracking problem of multiple-unmanned aerial vehicles (multi-UAVs). The problem is solved by decomposing into three sub-problems: information fusion, task assignment, and multi-UAV behavior decision-making.	
Specifically, in the information fusion process, we use the maximum consistency protocol to update the joint estimation states of multi-targets (JESMT) and the area detection information. The area detection information is represented by the equivalent visiting time map (EVTM), which is built based on the detection probability and the actual visiting time of the area.
Then, we model the task assignment problem of multi-UAV searching and tracking multi-targets as a network flow model with upper and lower flow bounds. An algorithm named task assignment minimum-cost maximum-flow (TAMM) is proposed.  
Cooperative behavior decision-making uses Fisher information as the mission reward to obtain the optimal tracking action of the UAV. Furthermore, a coverage behavior decision-making algorithm based on the anti-flocking method is designed for those UAVs assigned the coverage task. 
Finally, a distributed multi-UAV cooperative area coverage and target tracking algorithm is designed, which integrates information fusion, task assignment, and behavioral decision-making. 
Numerical and hardware-in-the-loop simulation results show that the proposed method can achieve persistent area coverage and cooperative target tracking.}




\maketitle

\section{Introduction}\label{sec1}
The multi-UAV cooperative area coverage and target tracking problem is a typical problem in area surveillance tasks, which is an essential means of information acquisition.
Typical application scenarios of area coverage and target tracking include forest fire monitoring, search and rescue, and pursuit-evasion. These applications not only require continuous search and coverage of the mission area. They also need the multi-UAV system to promptly detect the intruding non-cooperative targets and cooperatively track the allocated targets.  
However, various factors, such as the dynamic environment, the target appearance probability, and the sensor performance, make the problem very complex.

Many existing works consider the area coverage or target tracking problem sololy \cite{bib1,bib2}.The integrated design of area coverage and target tracking has not gotten enough attention \cite{bib1}. Previous studies of area coverage and target tracking problem usually include two categories: coupling or decouping. 
The coupling study of area coverage and target tracking problem simultaneously considers the coverage and tracking tasks, usually called simultaneous coverage and tracking (SCAT).
SCAT focuses on the multi-objective optimization of coverage and tracking tasks, and the typical method of SCAT is the voronoi-based coverage method \cite{bib2,bib3}.
The target tracking part in \cite{bib2} is treated as a parameter-based procedure. Then, the area coverage and target tracking problem is translated to the problem of covering environments with time-varying density functions. 
Although SCAT has been verified in actual scenarios, it ignores the single performance of area coverage or target tracking task to a certain extent.
For the decoupling design of coverage and tracking, it is necessary to make the UAV switch between different modes according to specific environments and mission requirements \cite{bib5} and reasonably allocate different UAVs to track multiple targets. Control logic design based on a finite state automaton model, integrating four modes of operations, is presented in \cite{bib8}. The semi-flocking algorithm in \cite{bib5,bib6,bib7} assigns a small flock of sensors to each target while at the same time leaving some sensors free to explore the environment. It enables mobile
nodes to self-organize themselves switch between searching and tracking modes. 
Although the above decoupling algorithms have been well applied,  they lack the coordinated management of UAVs, making it difficult to achieve optimal cooperation between UAVs.

Considering that the coupled algorithms ignores the performance of a single task, this paper solves the area coverage and target tracking problem under the decoupled framework. A hierarchical modular architecture is designed to enhance the collaboration of UAVs in the distributed manner. 
The hierarchical modular method integrates three sub-modules of information fusion, task allocation, and behavior decision-making. 




The information fusion module includes data preprocessing and information fusion methods.
Most of the current work is based on Bayesian methods \cite{bib24,bib25} for data pre-processing, while the sensors are modeled as sources of uncertainty.
While for the data fusion in a distributed system, nodes need to be guided through an interactive protocol to generate a consistent estimate. 
The information consensus algorithm \cite{bib28} has been widely used in distributed estimation \cite{bib29}, task assignment \cite{bib32,bib33}, and other scenarios. However, the consensus process is highly dependent on the communication links, and the algorithm convergence time increases rapidly as the task complexity increases. \cite{bib34} gives the termination condition of the maximum consensus algorithm in the distributed information fusion process. In this paper, we improve the maximum consensus information fusion algorithm in \cite{bib34} so that it can be applied to the area coverage and target tracking task.

The task allocation module is the key to enhancing the collaboration of UAVs. It can be solved by centralized or distributed methods. 
Centralized assignment methods \cite{bib36} can globally coordinate the complex relationships between tasks, but they rely too much on the control center and lack robustness.
Commonly used distributed allocation methods are based on the market mechanism \cite{bib44,bib45}.  
Distributed task assignment methods are computationally flexible and suitable for solving large-scale assignment problems. However, there are challenges to obtaining global optimal assignments.
In this paper, based on consensus algorithms and network flow theory \cite{bib49}, we design an algorithm that can get the global optimal allocation in a distributed manner. 
Applications of network flow theory to the task assignment are not typical. We find that the multi-UAV task assignment problem in area coverage and target tracking task can essentially be modeled as a minimum-cost maximum-flow problem \cite{bib50}, which then can be solved using various MCMF algorithms. 
The MCMF algorithms are centralized methods. Combining the MCMF algorithms with the consensus algorithms enables the allocation algorithm to adapt to the overall distributed architecture.

The collaborative behavioral decision-making module utilizes Fisher information
as the task reward, and UAVs make decisions according to the assigned coverage or tracking tasks.
As for coverage decision, researchers have focused on geometric, probabilistic or biological intelligence approaches \cite{bib10}.
Geometric-based methods \cite{bib13,bib14} can achieve complete coverage of the area.However, the centralized and offline nature makes the geometry-based approach unsuitable for dynamic environments.
The probabilistic-based approach \cite{bib10} builds probabilistic models to characterize the environmental uncertainty and uses search algorithms for distributed online decisions to reduce environmental uncertainty.
Compared with other decision-making algorithms, the methods based on biological intelligence can generate coverage behavior through simple rules with low computational cost \cite{bib19}. 
Therefore, we choose a rule-based anti-flocking algorithm \cite{bib22,bib23} for the coverage decision.
In the tracking behavior decision-making part, most target tracking methods are based on the principles of placing the target in the center of the field of view \cite{bib54}. In this paper, we use the rolling horizon method for tracking decision-making. And the optimal action sequence that maximizes the cumulative Fisher information volume is found.

According to the hierarchical modular design and the ideas of the three sub-modules, our main contributions are as follows:
\begin{itemize}
	\item First, a distributed hierarchical modular architecture 
	is provided for the area coverage and target tracking problem.
	We modularize information fusion, task assignment, and behavioral decision-making and integrate the three modules through a distributed architecture design.
	Compared with the coupling SCAT architecture, the distributed hierarchical modular design can fully exploit the capabilities of each sub-module, thereby improving the area coverage and target tracking performance of the system.
	
	\item Second, a distributed information fusion strategy that combines the compression, extraction, and fusion of the coverage time maps in the coverage task with the fusion of the target states is designed. The compression and extraction of the coverage information map allow our information fusion strategy to be adapted in arbitrarily large mission areas.
	
	\item Third, model the allocation problem of multi-UAV coverage and target tracking task as a network flow model.
	Then, a minimum-cost maximum-flow algorithm for task assignment is designed. 
	The allocation algorithm adapts to the change in the quantitative relationship between UAVs and targets and can achieve fast allocation for large-scale tasks.
\end{itemize}

The rest of the paper is organized as follows.  Section~\ref{sec2} analyzes the area coverage and target tracking problem and decomposes it into three subproblems: distributed information fusion, multi-UAV task assignment, and behavioral decision-making. And then, each subproblem is studied accordingly in Sections~\ref{sec3},\ref{sec4} and \ref{sec5}. Section~\ref{sec6} systematically designs a distributed multi-UAV cooperative area coverage and target tracking algorithm. Finally,  Section~\ref{sec7} gives the numerical and hardware-in-loop simulations and corresponding discussions. Numerical and hardware-in-the-loop simulations demonstrate that our proposed hierarchical modular architecture can effectively solve the area coverage and target tracking problem.

\section{Problem formulation}\label{sec2}
\begin{figure}[htbp]%
	\centering
	\includegraphics[width=0.6\textwidth]{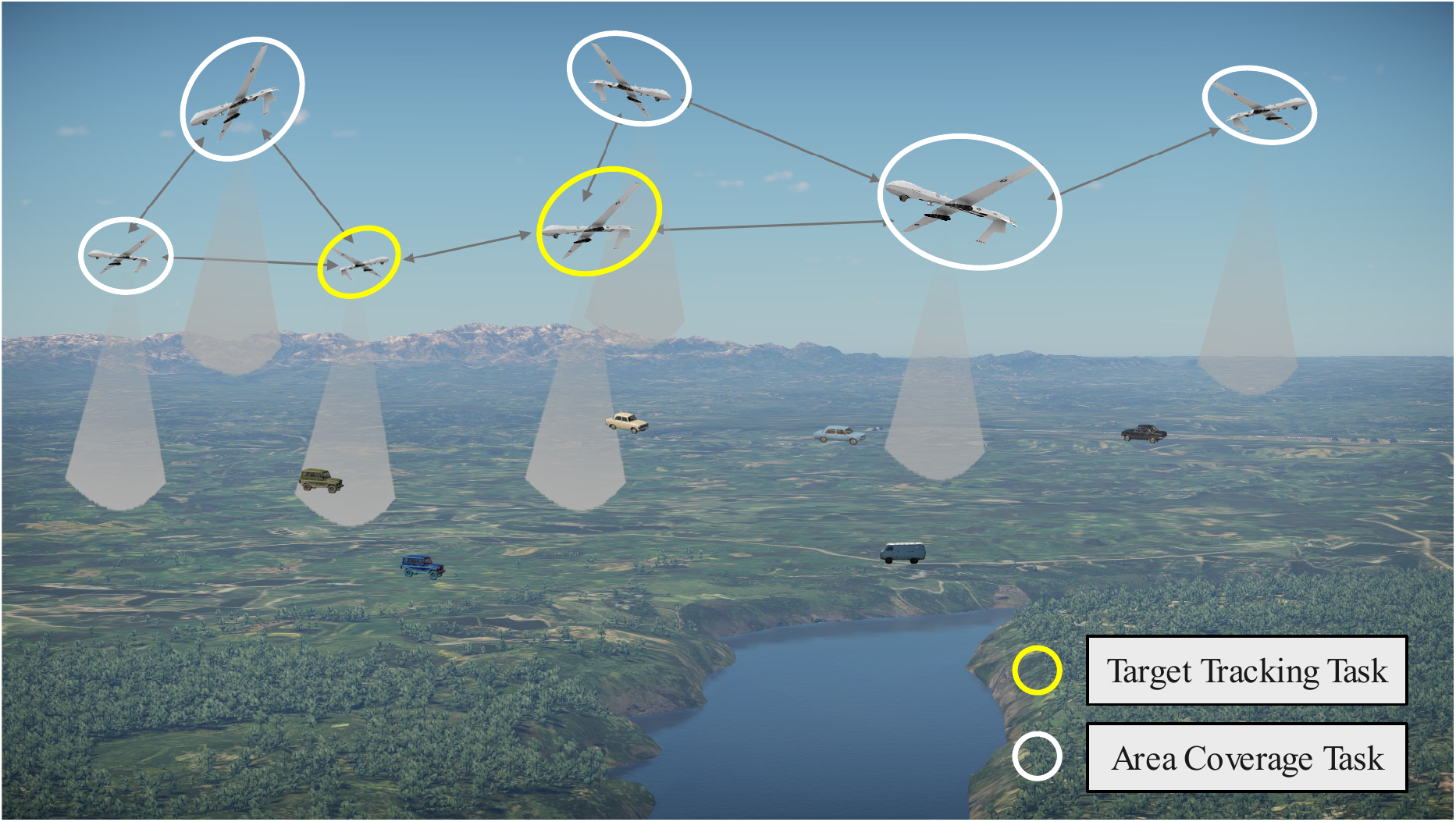}
	\caption{A typical scenario of the area coverage and target tracking task with multi-UAV system.}\label{fig1}
\end{figure}

In this paper, several ground targets are moving in the mission area and the number and states of the targets are unknown. 
The multi-UAV system consisting of $N_u$ homogeneous UAVs needs to cover the area continuously and automatically assigns UAVs to track the searched targets, while the remaining UAVs perform the coverage task. 
According to the mission requirements, each UAV has two task modes, i.e., the area coverage mode and the target tracking mode.
Fig.~\ref{fig1} shows a typical scenario of the area coverage and target tracking task.

Assuming that there are $N_\tau$ targets searched at the current moment. Let $I_{u}=\{1,\cdots,N_u\}$ denote the number list of all UAVs and $I_{\tau}=\{1,\cdots,N_\tau\}$ denote the number list of all targets. Beaides, ${\mathcal U}_i$ and ${\mathcal T}_j$ denote the $i$th UAV and $j$th target, respectively. 
The detection sensor is modeled by a disk with limited sensing capability and a fixed field of view (FOV). 
While the $i$th UAV can observe any targets within its measurement range $r_{i}^o$, which is determined by its altitude $h_i$ and FOV.
UAVs communicate through the wireless communication module with the communication range of $r_c$. Let $p_{i}=\left[x_{i},y_{i}\right]$ denote the position of ${\mathcal U}_{i}$. Then we use $\mathscr{G}_i^n=\{\mathscr{V}_i^n,\mathscr{E}_i^n\}$ to represent the undirected communication subgraph containing ${\mathcal U}_i$ and its neighbors, where $\mathscr{V}_i^{n}=\{ q\in I_u \mid \left \| p_{q}-p_{i} \right \| < r_{c}\}$ and $\mathscr{E}_i^n = \{(i,q) \mid q\in\mathscr{V}_i^n,q\neq i\}$. Here $\mathscr{G}_i^n$ depends on the communication range $r_c$ and the relative location between UAVs. 

\begin{figure}[htbp]%
	\centering
	\includegraphics[width=0.4\textwidth]{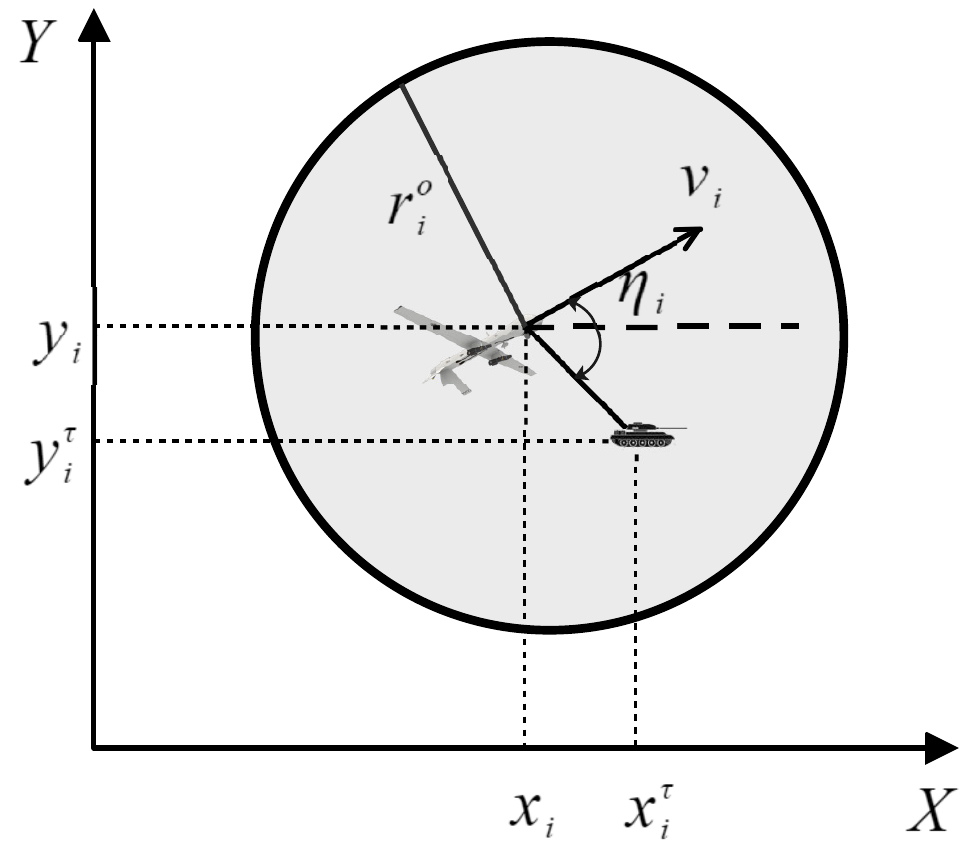}
	\caption{The geometric relationship of UAV and target in the two-dimensional plane.}\label{fig2_UT}
\end{figure}

Let $s_{j}^{\tau}=(x_{j}^{\tau},\dot{x}_{j}^{\tau},y_{j}^{\tau},\dot{y}_{j}^{\tau})$ represent the state of target ${\mathcal T}_j$. We use the linear kinematic model for the ground moving targets. Meanwhile, we assume that the UAVs all fly at a constant altitude. The state of ${\mathcal U}_{i}$ is defined by $s_{i}=\left( x_{i},y_{i}, v_{i},{\eta}_{i}  \right)$, where $v_{i}$ is the flight speed and ${\eta}_{i}$ is the heading angle. The geometric relationship of UAV and target in the two-dimensional plane is shown in Fig.~\ref{fig2_UT}. The discrete kinematics model of the fixed-wing UAV is as follows:
\begin{equation}
	\left\{
	\begin{aligned}
		x_{i}\left(k+1\right) &= x_{i}\left(k\right) + v_{i}\left(k\right) \Delta T\cos{\eta}_{i}\\
		y_{i}\left(k+1\right) &= y_{i}\left(k\right) + v_{i}\left(k\right) \Delta T\sin{\eta}_{i}\\
		v_{i}\left(k+1\right) &=\left[v_{i}\left(k\right) + \Delta v_{i}\left(k\right) \Delta T \right]_{v_{min}}^{v_{max}}\\
		{\eta}_{i}\left(k+1\right) &= {\eta}_{i}\left(k\right) + {\omega}_{i}\left(k\right)\Delta T
	\end{aligned}
	\right. 
\end{equation}
where $\Delta T$ denotes the length of each time step, and the action command $\pi_{i}$ of the $i$th UAV is $\left(\Delta v_{i},{\omega}_{i}\right)$, indicating the ground acceleration and heading angular velocity.

In our work, we only consider the target tracking task in the task allocation stage, since moving targets are more valuable.
We calculate the tracking action and the corresponding reward of each UAV to each target through the tracking decision algorithm and then assign tasks to UAVs according to the reward matrix $R \in  {\mathbb{R}}^{N_u \times N_\tau}$. $R_{ij}$ is the corresponding task reward when ${\mathcal U}_i$ tracks target ${\mathcal T}_j$. 
 $\digamma \in  {\mathbb{R}}^{N_u \times N_\tau}$ denotes the task decision matrix. $\digamma_{ij}=1$ indicates that task ${\mathcal T}_j$ is assigned to ${\mathcal U}_i$ and 0 otherwise. $\pi \in  {\mathbb{R}}^{N_u}$ denotes the action command of all UAVs. And $\pi_{i}$ is the action command of ${\mathcal U}_i$.
Our goal is to find the optimal mapping $\digamma$ from the tasks to UAVs and plan the action of each UAV to maximize the overall mission reward while meeting the task constraints. Then, the optimization function of the multi-UAV system is:
\begin{equation}
	\begin{aligned}
		& \max \limits_{\digamma,\pi} \sum\limits_{i=1}^{N_u} {\sum\limits_{j=1}^{N_\tau}{R_{ij}(\pi_{i})\digamma_{ij}} }\\
		& \begin{array}{r@{\quad}r@{}l@{\quad}l}
		s.t.
		 &\sum\limits_{i=1}^{N_u} \digamma_{ij}&\leq {n_j},    &\forall{j} \in I_{\tau}\\
		 &\sum\limits_{j=1}^{N_\tau} \digamma_{ij}&\leq 1,    &\forall{i} \in I_{u}\\
		 &\digamma_{ij}\in\{&0,1\},    &\forall{i} \in I_{u}, \forall{j} \in I_{\tau}\\
		\end{array}
	\end{aligned}
	\label{eq1}
\end{equation}
The first constraint indicates that the number of UAVs tracking one target cannot exceed $n_j$. While the second constraint shows that each UAV can select at most one target to track.
Note that if a UAV is assigned no target to track, it then performs the area coverage tasks. 

\begin{figure}[htbp]%
	\centering
	\includegraphics[width=0.8\textwidth]{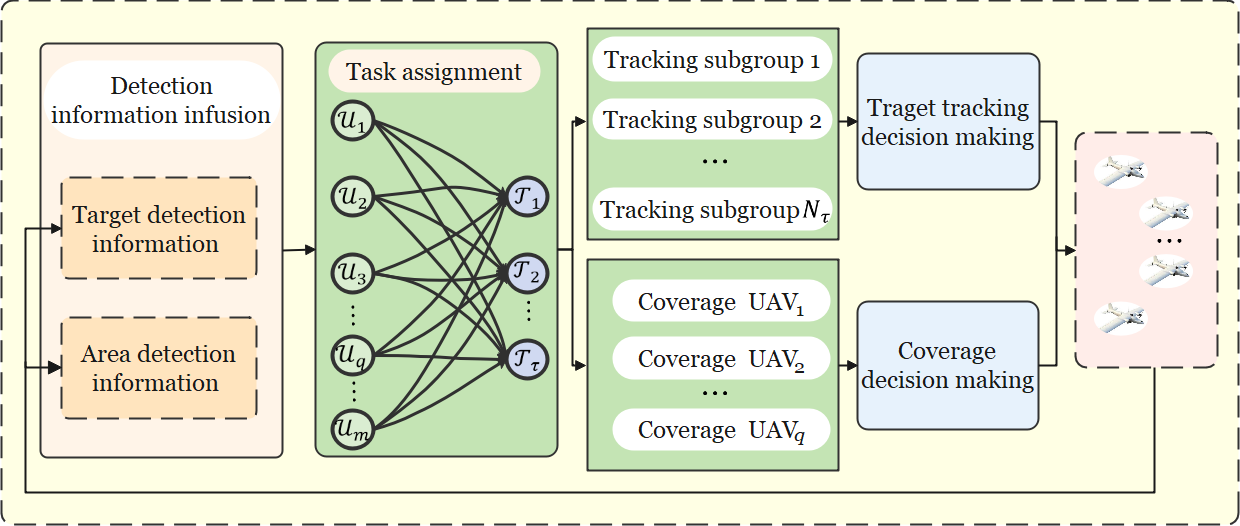}
	\caption{The overall framework of the multi-UAV area coverage and target tracking system.}\label{fig2}
\end{figure}
In the decoupling architecture, we divide the area coverage and target tracking task into three sub-problems: information fusion, task assignment, and multi-UAV behavior decision-making. Fig.~\ref{fig2} is the framework of our multi-UAV area coverage and target tracking system. Then we analyze each sub-problem in detail in the following sections.

\section{Multi-UAV information fusion}\label{sec3}
The information interaction is essential for realizing the cooperation among UAVs. It can improve the area coverage efficiency and the tracking accuracy of targets. In this section, a distributed approach is proposed where the joint estimation state of multi-targets and the coverage information of each UAV are propagated through the whole network so that the consistency of the detection information is guaranteed.  The information fusion algorithm in this section improves the information fusion strategy in \cite{bib34} which is designed only for the target tracking task.

\subsection{Definition of detection information}\label{subsec2}
In the information fusion process, UAVs need to exchange their detection information (the local area coverage information and the local estimation state of multi-targets) with neighbors. Therefore,  we design the area information map  and a target information table to save the detection information of each UAV.

\subsubsection{Equivalent visiting time map}\label{subsubsec2}

In order to record the coverage information of each UAV and cooperate with other UAVs, we discretize the task area into a grid map $g$ with $M*N$ grids. Suppose we set the detection radius of the airborne sensor simply as a disk. The target will be covered as long as within the sensor's detection range, which does not consider the probability of successful detection. Therefore, we build an equivalent visiting time map (EVTM) considering the sensor's performance.

Let ${T}_{i} \in {\mathbb{R}}^{M \times N}$ denote the EVTM of the $i$th UAV. The equivalent visiting time map (See Fig.~\ref{fig3_EVTM}) records the equivalent time when the grids were last visited. 
${t}_{i}^{m,n}(k-1)$ represents the equivalent visiting time of grid $g \left(m,n\right)$ at the end of time step $k-1$. 
$t_{k}$ is the actual time at time step $k$.

Initialize the equivalent visiting time map with ${t}_{i}^{m,n} \left(0\right)=t_0$, and the update rule of the EVTM is as follows:
\begin{equation}
	\hat{t}_{i}^{m,n} ({k})=\left\{
	\begin{aligned}
		& {t}_{i}^{m,n}(k-1),    &g(m,n)\ not\ visited\ at\ t_k\\
		& t_{k} - \Delta t_{i}^{m,n}(k),    &g(m,n)\ visited\ at\ t_k
	\end{aligned}
	\right. 
	\label{eq3}
\end{equation}
$\Delta t_{i}^{m,n}(k)$ is relevant to the probability of successful detection and will be calculated subsequently. $\hat{t}_{i}^{m,n} ({k})$ is updated only by the detection information of ${\mathcal U}_i$. After information fusion phase, we can get ${t}_{i}^{m,n} ({k})$ with the dection information of the UAVs in the same connected network.

Let $\gamma_{i}^d(k,m,n)$ denote the successful detection probability of  grid $g(m,n)$ by ${\mathcal U}_i$ at time step $k$. $\gamma_{i}^d(k,m,n)$ is related to the sensor performance and the distance $dis_{i\rightarrow (m,n)}$ between ${\mathcal U}_i$ and the detection grid $g(m,n)$, that is:
\begin{equation}
	\begin{aligned}
		{\gamma_{i}^d} \left(m,n\right) = e^{\left[-w_{p1} 
			\left(dis_{i\rightarrow \left(m,n\right)}/w_{p2} \right)^{w_{p3}}
			\right]}
	\end{aligned}
\end{equation}
where $w_{p1}$, $w_{p2}$, and $w_{p3}$ are adjustable parameters. 

Considering the successful detection probability $\gamma_{i}^d(k,m,n)$, the visiting requirement of grid $g(m,n)$ is:
\begin{equation}
	\begin{aligned}
		{\lambda}_{i}^{m,n} ({k}) = {\lambda}_{i}^{m,n}(k-1) \left(1- \gamma_{i}^d(k-1,m,n) \right)
	\end{aligned}
\end{equation}

We can also use an S-curve function to represent the visiting requirement ${\lambda}_{i}^{m,n} (k)$ of grid $g(m,n)$ from its corresponding equivalent visiting time:
\begin{equation}
	\begin{aligned}
		{\lambda}_{i}^{m,n} (k) = 1 - e^{-\alpha \left[\left(t_{k}-{t}_{i}^{m,n} (k) \right) /{T_{c}} \right]^\beta}
	\end{aligned}
	\label{eq6}
\end{equation}
where $\alpha$ and $\beta$ are curve parameters, $T_{c}$  is the revisit time threshold.

According to Eq.~\ref{eq3}-\ref{eq6}, we have:
\begin{small}
	\begin{equation}
		\begin{aligned}
			\Delta t_{i}^{m,n} (k)= & {T_{c}} \left[-\frac{1}{\alpha} ln\left[1\!-\! \left(1\!-\!e^{\!-\!\alpha \left[\left(t_{k-1}\!-\! {t}_{i}^{m,n} \left(k-1\right) \right) /{T_{c}} \right]^\beta} \right) \right.\right.	\\
			& \left.\left. \left(1\!-\! \gamma_{i}^d (k-1,m,n) \right)
			\right] \right]^{1/\beta}
		\end{aligned}
	\end{equation}
\end{small}

\subsubsection{Target information table}\label{subsubsec3}
In our previous work \cite{bib34}, each UAV maintains an information table to store its current estimation states of the targets. We define the information for a single target at time step $k$ as the following tuple:
\begin{equation}
	\{ \hat{s}_{i,j}(k), \hat{P}_{i,j}(k), \hat{\rho}_{i,j}(k), \overline{s}_{i,j}(k), \overline{P}_{i,j}(k)\}
\end{equation}
$\hat{s}_{i,j}$ and $\hat{P}_{i,j}$ are the local estimated state and the error covariance matrix of target ${\mathcal T}_j$
and can be obtained  through the local Kalman filtering of ${\mathcal U}_i$.
 Perceptual confidence $\hat{\rho}_{i,j}$ quantifies the accuracy of the UAV's target state estimation and is set as the trace of the covariance matrix, which is:
\begin{equation}
	 \hat{\rho}_{i,j}(k):=\{Trace (\hat{P}_{i,j}(k))\}^{-1}
\end{equation}
Besides, $\overline{s}_{i,j}$ and  $\overline{P}_{i,j}$ are the corresponding estimated state and error covariance matrix after information fusion. Therefore, we can use $\{ \{ \hat{s}_{i,j}, \hat{P}_{i,j}, \hat{\rho}_{i,j} \} \mid j \in I_\tau \}$ to denote the local detection information of ${\mathcal U}_i$ to all targets.

\subsection{Distributed information fusion}\label{subsec3}
In the distributed information fusion process, each UAV exchanges information with its neighbors to update the local information.
The coverage time map, the perceptual confidence value, and the corresponding estimated state and error covariance matrix are propagated over the network in a finite time.  Through limited updates with max-consensus protocol, the  perceptual information of all UAVs in the connected network achieves consistency.

\subsubsection{Communication topology}

\begin{figure}[htbp]%
	\centering
	\subfigure[]{
		\begin{minipage}[c]{0.30\linewidth}
			\centering
			\includegraphics[width = 1.3in,height=1in]{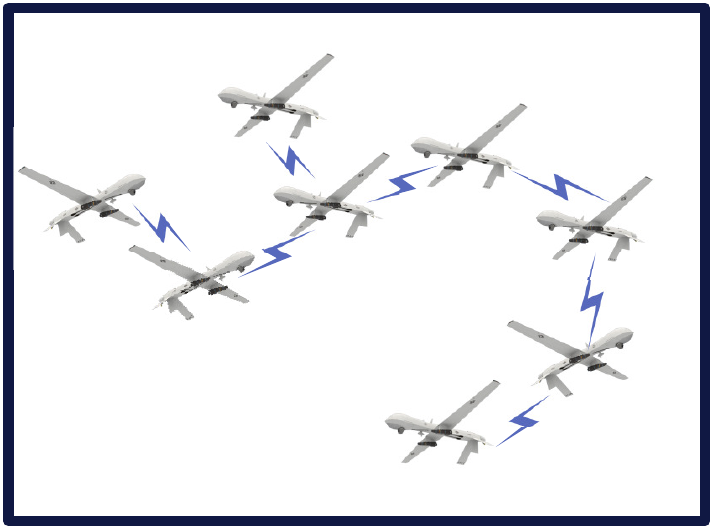}
			\label{fig4Tya}
		\end{minipage}%
	}%
	\subfigure[]{
		\begin{minipage}[c]{0.3\linewidth}
			\centering
			\includegraphics[width = 1.3in,height=1in]{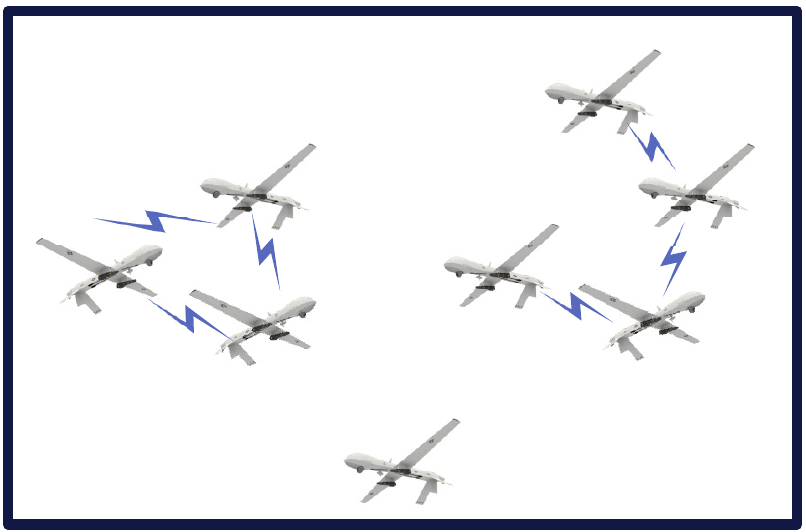}
			\label{fig4Tyb}
		\end{minipage}%
	}%
	\subfigure[]{
		\begin{minipage}[c]{0.3\linewidth}
			\centering
			\includegraphics[width = 1.3in,height=1in]{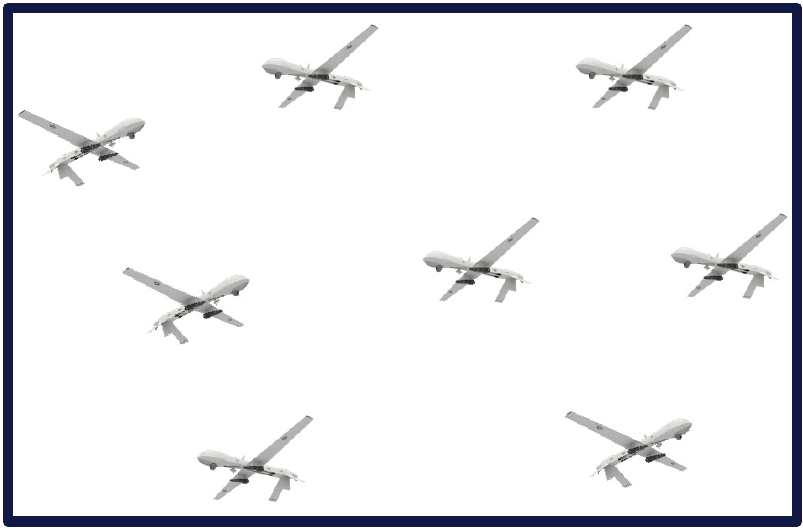}
			\label{fig4Tyc}
		\end{minipage}%
	}%
	\caption{The communication topologies of the multi-UAV system.}\label{fig4_Ty}
\end{figure}

When performing the area coverage and target tracking task, UAVs are constantly moving. Thus the communication topology of the multi-UAV system is time-varying. 
Depending on the spatial distribution and communication radius of UAVs, several communication topologies may emerge as shown in Fig.~\ref{fig4_Ty}.
\begin{enumerate}[(a)]
	\item {The connected communication topology;}
	
	\item {The communication topology which is divided into several isolated connected communication topologies;}
	
	\item {The communication topology that all nodes can not communicate with each other.}
\end{enumerate}

To simplify the analysis, the communication topology is assumed to remain constant during each time step. Only the UAVs within the same connected communication topology  need to exchange and fuse information to make the detection information achieve consistency.

\subsubsection{The compression and extraction of EVTM}
To reduce the interactive information, we compress and extract the visiting time map to get a compressed time map and a local time map, representing the global coverage information and local coverage information, respectively.
\begin{figure}[htbp]%
	\centering
	\includegraphics[width=0.6\textwidth]{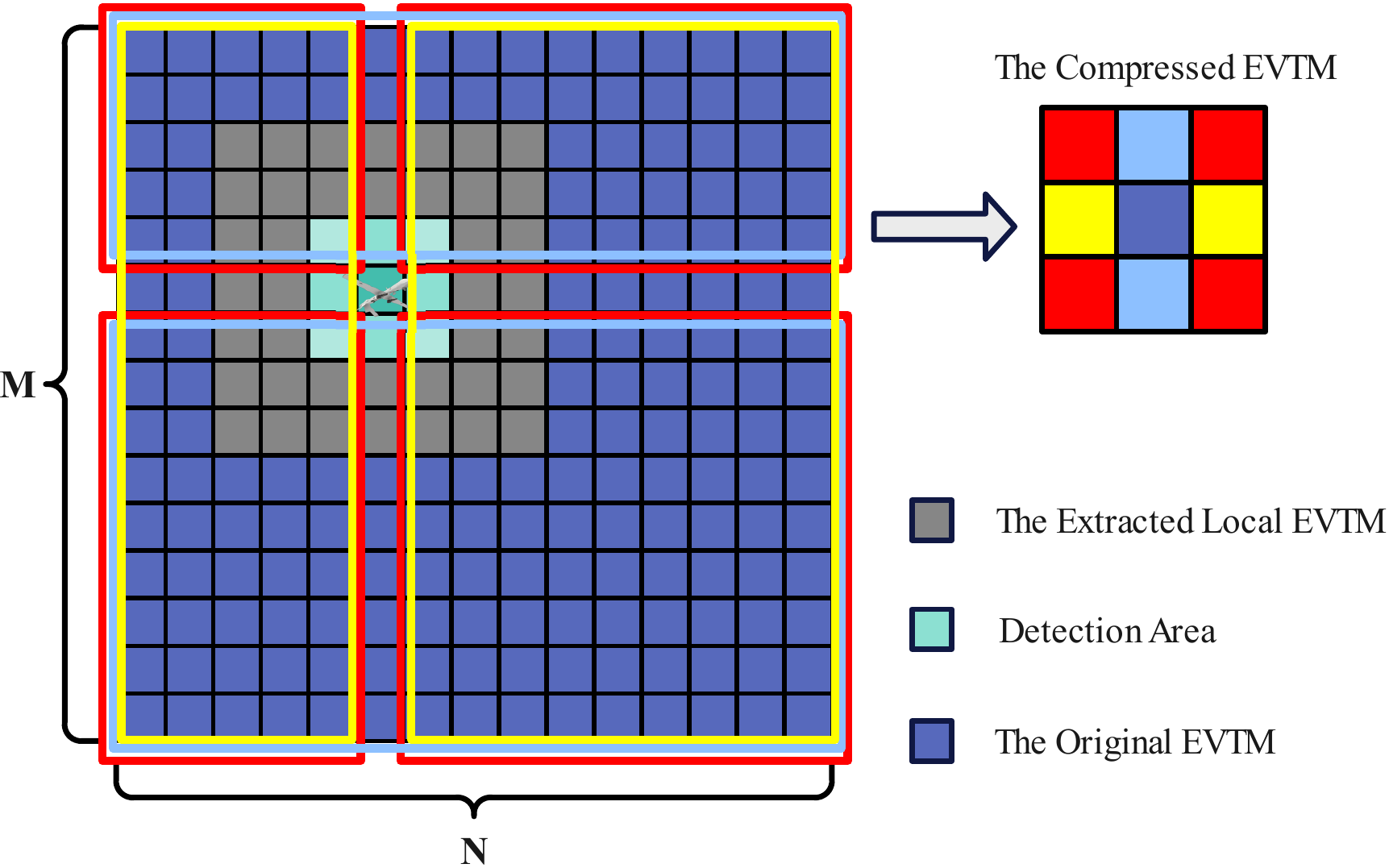}
	\caption{ The schematic of the compression and extraction of EVTM.}\label{fig3_EVTM}
\end{figure}

Fig.~\ref{fig3_EVTM} shows the schematic of the compression and extraction of the EVTM. Assuming that ${\mathcal U}_i$ is located at grid $g(m_i,n_i)$, the compressed time map $\hat{T}_{i}^{g} \in {\mathbb{R}}^{3 \times 3}$ can be obtained by averaging the equivalent visiting time of the grids around $g(m_i,n_i)$ in EVTM $\hat{T}_{i}$, which is:
\begin{small}
	\begin{equation}
		\begin{split}
			\hat{T}_{i}^{g} &= compress(\hat{T}_{i},(m_i, n_i))\\
			&= \left(
			\begin{array}{ccc}
				\frac{\sum\nolimits_{m_{i}+1}^{M}\sum\nolimits_{1}^{n_{i}-1} {\hat t}_{i}^{m',n'}}{(M-m_i)(n_{i}-1)}
				&\frac{\sum\nolimits_{m_{i}+1}^{M}\sum\nolimits_{1}^{N} {\hat t}_{i}^{m',n'}}{(M-m_i)N} 
				&\frac{\sum\nolimits_{m_{i}+1}^{M}\sum\nolimits_{n_{i}+1}^{N} {\hat t}_{i}^{m',n'}}{(M-m_i)(N-n_{i})}  \\
				\frac{\sum\nolimits_{1}^{M}\sum\nolimits_{1}^{n_{i}-1} {\hat t}_{i}^{m',n'}}{M(n_{i}-1)}
				&{\hat t}_{i}^{m_{i},n_{i}}
				&\frac{\sum\nolimits_{1}^{M}\sum\nolimits_{n_{i}+1}^{N} {\hat t}_{j}^{m',n'}}{M(N-n_{i})}  \\
				\frac{\sum\nolimits_{1}^{m_{i}-1}\sum\nolimits_{1}^{n_{i}-1} {\hat t}_{i}^{m',n'}}{(m_{i}-1)(n_{i}-1)}
				&\frac{\sum\nolimits_{1}^{m_{i}-1}\sum\nolimits_{1}^{N} {\hat t}_{i}^{m',n'}}{(m_{i}-1)N} 
				&\frac{\sum\nolimits_{1}^{m_{i}-1}\sum\nolimits_{n_{i}+1}^{N} {\hat t}_{j}^{m',n'}}{(m_{i}-1)(N-n_{i})}  
			\end{array}
			\right)
		\end{split}
	\end{equation}
\end{small}
where $compress(\cdot)$ is the compression function of the equivalent visiting time map. 

The local equivalent time map  $\hat{T}_{i}^l \in {\mathbb{R}}^{L \times L}$ is obtained by extracting the $L\times L$ grids around the UAV's location $g(m_i,n_i)$ from the EVTM $\hat{T}_{i}$. The extraction function $extract(\cdot)$ is as follows:
\begin{small}
\begin{equation}
	\begin{aligned}
	\hat{T}_{i}^l &= extract(\hat{T}_{i},(m_i, n_i))\\
	&= \{
	\hat{t}_{i}^{m,n} \mid m \in \{m_i,m_i \pm (L-1)/2\}, n \in \{n_i,n_i \pm (L-1)/2\}
	\}
	\end{aligned}
    \label{eq11}
\end{equation}
\end{small}
$L$ is a positive odd number.

Table~\ref{tab_0} shows the message passed from neighboring UAV ${\mathcal U}_q$ to ${\mathcal U}_i$, of which  $\hat{T}_{q}^l$ is the local equivalent time map of ${\mathcal U}_q$.
Such information interaction allows our coverage algorithm to extend to surveillance tasks with arbitrarily large mission areas.

\begin{table}[h!]
	\begin{center}
    \begin{minipage}{0.75\textwidth}
	\caption{Information Interaction} 
	 \footnotesize
	 \renewcommand{\arraystretch}{0.8}
	\begin{tabular}{@{}cc@{}} 
		\toprule 
		\textbf{Information type} & \textbf{Information content} \\ 
		\midrule 
		Status information & $s_{q}(k)=(x_{q}(k),y_{q}(k),v_{q}(k), \eta_{q}(k))$ \\   \midrule 
		Area detection information & $ \hat{T}_{q}^{l}(k)$ \\
		\midrule 
		Target detection information & $\{ \{ \hat{s}_{q,j}(k), \hat{P}_{q,j}(k), \hat{\rho}_{q,j}(k) \} \mid j \in I_\tau \}$ \\ 
		\botrule
	\end{tabular}
	\label{tab_0} 
\end{minipage}
\end{center}
\end{table}

\begin{remark}
	Regardless of the size of the area, the interaction information between UAVs is always a map with $L \times L$ grids. 
	Then, for an arbitrarily large detection area with an area information map of $M \times N$ grids,  the information content is only $(L\times L)/(M \times N)$ of the original global map.
\end{remark}

\subsubsection{The distributed information fusion based on maximum consistency}
In order to get consistent information before the subsequent decision-making process, we define a new sampling time denoted by index $d$  with a higher frequency in the information fusion phase. Moreover, to express the fusion process more clearly, a group of temporary variables initialized with the information at time step $k$ (See Table~\ref{tab_1}) is defined. 
\begin{table}[h!]
	\begin{center}
		\begin{minipage}{1\textwidth}
			\caption{Temporary variables} 
			\footnotesize
			\renewcommand{\arraystretch}{0.8}
			\begin{tabular}{@{}cc@{}} 
				\toprule 
				\textbf{Temporary variable} & \textbf{Initialization} \\ 
				\midrule 
				$\hat{T'}_{q}(d)$ & $\hat{T'}_{q}(0) = \hat{T}_{q}(k)$ \\
				\midrule 
				$\hat{T'}_{q}^l(d)$ & $\hat{T'}_{q}^l(0) = extract(\hat{T'}_{q}(0),(m_q,n_q))$ \\
				\midrule 
				$\{ \hat{s'}_{q,j}(d), \hat{P'}_{q,j}(d), \hat{\rho'}_{q,j}(d) \}_ {j\in I_\tau} $ & $ \{ \hat{s'}_{q,j}(0), \hat{P'}_{q,j}({0}), \hat{\rho'}_{q,j}({0}) \} = \{  \hat{s}_{q,j}(k), \hat{P}_{q,j}(k), \hat{\rho}_{q,j}(k) \},\forall{j} \in I_\tau $ \\ 
				\botrule
			\end{tabular}
			\label{tab_1} 
		\end{minipage}
	\end{center}
\end{table}

The detection information from neighbors can then be fused based on the max-consensus protocol. 
Specifically, the equivalent visiting time map $\hat{T'}_i(d)$ is updated with the received EVTM as follows:
\begin{equation}
	\hat{t'}_i^{m,n}(d) = \max \limits_{u \in \mathscr{V}_i^n} \{ \hat{t'}_u^{m,n}(d-1) \}
	\label{eq12}
\end{equation}
The update rule of the  perceptual confidence is:
\begin{equation}
	{\hat{\rho}'}_{i,j}(d) = \max \limits_{u \in \mathscr{V}_i^n} \{{\hat{\rho}'}_{u,j}({d-1}) \}, j \in I_\tau
	\label{eq13}
\end{equation}

According to the general definition of maximum consistency in \cite{bib52}, we give the concept of maximum consistency of the detection information in our area coverage and target tracking system.
\begin{definition}[Maximum consistency of the detection information]
	Consider the undirected communication subgraph $\mathscr{G}_i=\{\mathscr{V}_i,\mathscr{E}_i\}$ where ${\mathcal U}_i$ is located, the initial value $\hat{t'}_u^{m,n}(0)$ of the equivalent visiting time map, and the initial value $\hat{\rho}'_{u,j}(0)$ of perceptual confidence of each UAV in the subgraph, 
	the update rules of EVTM and the perceptual confidence are (\ref{eq12}) and (\ref{eq13}), respectively. Then the system achieves maximum consistency, if $\exists \delta \in \mathbb{N}$, so that:
	\begin{align} 
		\hat{\rho}'_{i,j}(d) & =  \hat{\rho}'_{q,j}(d) \notag \\
		& = \max \limits_{u \in \mathscr{V}_i} \{{\hat{\rho}'}_{u,j}(0) \},{\forall j} \in I_\tau, {\forall d} \geq \delta, \forall i,q \in \mathscr{V}_i \\
		\hat{t'}_i^{m,n}(d) & =  \hat{t'}_q^{m,n}(d) \notag \\
		& = \max \limits_{u \in \mathscr{V}_i} \{ \hat{t'}_u^{m,n}(0) \}, {\forall d} \geq \delta, \forall i,q \in \mathscr{V}_i, \forall m  \leq M, \forall n \leq N 
	\end{align}
	where $\delta$ is the lower bound of the iteration number to achieve maximum consistency in the whole connected graph $\mathscr{G}_i$. 
\end{definition}

According to the termination condition of the maximum consistency algorithm given in Theorem 1 of \cite{bib34}, we directly set the number of iterations of the $i$th UAV as ${\mathcal D}_i$, which is the diameter of the shortest path tree (SPT) of $\mathscr{G}_i$ rooted at node $i$. Algorithm~\ref{algo_1} gives the information fusion process of the $i$th UAV. After the information fusion phase, we get the joint estimation states of multi-targets (JESMT) and the joint EVTM of the mission area.

\renewcommand{\algorithmicrequire}{\textbf{Input:}}
\renewcommand{\algorithmicensure}{\textbf{Output:}}
\begin{algorithm}
	\caption{Information fusion of the $i$th UAV at the time step $k$}\label{algo_1}
	\begin{algorithmic}[1]
		\Require ${T}_{i}(k-1), \{ \{ \overline{s}_{i,j}({k-1}), \overline{P}_{i,j}({k-1}) \} \mid j \in I_\tau \}   $ 
		\Ensure $ {T}_{i}(k), {T}_{i}^l(k),{T}_{i}^g(k), \{ \{ \overline{s}_{i,j}({k}), \overline{P}_{i,j}({k}) \} \mid j \in I_\tau \}  $
		\For{$j = 1$ to $N_\tau$} 
		\State Calculate $\{ \hat{s}_{i,j}({k}), \hat{P}_{i,j}({k}) \}$ by Kalman filter according to $\{ \overline{s}_{i,j}({k-1}), \overline{P}_{i,j}({k-1}) \}$;
		\State $\hat{\rho}_{i,j}({k}) := \{Trace (\hat{P}_{i,j}({k}))\}^{-1}$;
		\EndFor
		\State Get $\hat{T}_i(k)$ from ${T}_{i}(k-1)$ by (\ref{eq3});
		\State Calculate  $\hat{T}_i^l({k})$ according to (\ref{eq11});
		\State Let $ \{ \hat{s'}_{i,j}({0}), \hat{P'}_{i,j}({0}), \hat{\rho'}_{i,j}({0}) \} = \{  \hat{s}_{i,j}({k}), \hat{P}_{i,j}({k}), \hat{\rho}_{i,j}({k}) \},\forall{j} \in I_\tau $,
		$\hat{T'}_i(0) = \hat{T}_i({k})$, and
		$\hat{T'}_{i}^l(0) = \hat{T}_i^l({k})$;
		\For{$d = 1$ to ${\mathcal D}_i$}
		\State Send $\{ \{ \hat{s'}_{i,j}({d-1}), \hat{P'}_{i,j}(d-1), \hat{\rho'}_{i,j}(d-1) \} \mid \forall{j} \in I_\tau \}$, $ \hat{T'}_i^l(d-1)$;
		\State Receive $\{ \{ \hat{s'}_{u,j}({d-1}), \hat{P'}_{u,j}({d-1}), \hat{\rho'}_{u,j}({d-1}) \} \mid \forall{j} \in I_\tau \}$, $ \hat{T'}_u^l(d-1)$, for $\forall{u} \in \mathscr{V}_i^n $;
		\State Update $\hat{T'}_{i}$ according to (\ref{eq12}):
		\State $q \gets \mathop{\arg\max}\limits_{u \in \mathscr{V}_i^n} {\hat{t'}_u^{m,n}(d-1)}$, $ \hat{t'}_i^{m,n}(d) = \hat{t'}_q^{m,n}(d-1)$;
		\For{$j=1$ to $N_\tau$}
		\State $l \gets \mathop{\arg\max}\limits_{u \in \mathscr{V}_i^n} {\hat{\rho'}}_{u,j}({d-1})$;
		\State $\{ \hat{s'}_{i,j}({d}), \hat{P'}_{i,j}({d}), \hat{\rho'}_{i,j}({d}) \} = \{ \hat{s'}_{l,j}({d-1}), \hat{P'}_{l,j}({d-1}), \hat{\rho'}_{l,j}({d-1}) \} $;
		\EndFor
		\State $\hat{T'}_i^l(d) = extract(\hat{T'}_i(d),(m_i,n_i))$;
		\EndFor
		\State ${T}_{i}(k) = \hat{T'}_{i}({\mathcal D}_i)$, ${T}_{i}^l(k) = \hat{T'}_{i}^l({\mathcal D}_i) $, ${T}_{i}^g(k) = compress(\hat{T}_{i}(k),(m_i,n_i))$, $\{ \overline{s}_{i,j}(k), \overline{P}_{i,j}(k) \} = \{ \hat{s'}_{i,j}({{\mathcal D}_i}), \hat{P'}_{i,j}({{\mathcal D}_i}) \}$, ${\forall j \in I_\tau}$.
	\end{algorithmic}
\end{algorithm}

\begin{remark}
	It is worthy pointing out that the information fusion strategy in \cite{bib34} can only be used for the target tracking task. 
	By contrast, Algorithm~\ref{algo_1} in this paper adds the compression, extraction, and fusion of the coverage time map for the area coverage and target tracking task.
\end{remark}
\section{Task assignment based on minimum-cost maximum-flow method}\label{sec4}
According to the task allocation problem of multi-UAV search and tracking multi-target formulated in (\ref{eq1}),  it is assumed that target ${\mathcal T}_j$ should be tracked by at least one UAV and at most $n_j$ UAVs. Meanwhile, each UAV can choose one target to track at most.
Combined with the network flow theory, this task allocation problem can be modeled as a network flow model with upper and lower flow bounds.

\subsection{Related concepts of network flow}\label{sec4_1}

We first give some definitions related to network flow problem \cite{bib49,bib50}. 

\begin{enumerate}[ 1) ]
	\item {Capacity Network: Given a directed graph $\mathcal G = (\mathcal V, \mathcal A)$ and the capacity $c(\nu_i,\nu_j)$ of each arc $a_{i,j}=(\nu_i,\nu_j)$, $(\mathcal V, \mathcal A, c(a_{i,j}))$ is called the capacity network.}
	
	\item {Source, Sink, and Intermediate Vertices: In a capacity network, the source node is denoted as $\nu_s$ with in-degree zero, the sink node $\nu_t$ has out-degree
		zero, and the other nodes are called intermediate vertices.}
	
	\item {Network Flow: A function $f: \mathcal A \to \mathcal R$ defined from the arc set $\mathcal A$ to the nonnegative number set is called the network flow on $\mathcal G$, while $f_{ij}=f(\nu_i,\nu_j)$ is the flow on arc $a_{ij}$.}
	
	\item {Feasible Flow: A feasible flow is a network flow $f$ from $\nu_s$ to $\nu_t$, which simultaneously satisfies the following capacity constraints (\ref{eq16}) and conservation constraints (\ref{eq17}) .
		\begin{equation}
			0 \leq f(\nu_i,\nu_j) \leq c(\nu_i,\nu_j), \quad {\forall (\nu_i,\nu_j) \in \mathcal A}
			\label{eq16}
		\end{equation}
		\begin{equation}
			\sum\nolimits_{(\nu_i,\nu_j) \in \mathcal A} {f_{ij}} = \sum\nolimits_{(\nu_j,\nu_i) \in \mathcal A} {f_{ji}} 
			, \quad \forall \nu_i \in \mathcal V \backslash \{\nu_s,\nu_t\}
			\label{eq17}
		\end{equation}
		While the value of a feasible flow is defined as:
		\begin{equation}
			\vert f \vert = \sum\nolimits_{\nu_j \in \mathcal V} {f_{sj}} = \sum\nolimits_{\nu_j \in \mathcal V} {f_{jt}}
			\label{eq18}
		\end{equation}
	}
	
	\item {Upper and Lower Bounds of Flow: For each arc $(\nu_i,\nu_j)$ of the directed graph $\mathcal G = (\mathcal V, \mathcal A)$, the capacity constraints (\ref{eq16}) can be further extended with the upper bound flow $upper_{ij}$ and the lower bound flow $lower_{ij}$.}
	
	\item {Minimum-cost maximum-flow: Given a capacity network $(\mathcal V, \mathcal A, c(a_{ij}))$, the cost of the unit flow for arc $a_{ij}$ is denoted as $l(\nu_i,\nu_j)$. The minimum-cost maximum-flow problem is to find a feasible flow $f: \mathcal A \to \mathcal R$, so that
		\begin{equation}
			\min { \sum_{(\nu_i,\nu_j) \in \mathcal A} l_{ij} f_{ij}  } \quad \& \quad
			\max{\vert f \vert}    
			\label{eq19}
		\end{equation}
	}
\end{enumerate}

\subsection{Network flow model for multi-UAV task assignment}\label{sec_4_2}
According to the concepts of network flow introduced in the previous section, the task assignment problem of multi-UAV searching and tracking multi-target can be modeled as a network flow model with upper and lower flow bounds.

\begin{figure}[htbp]%
	\centering
	\subfigure[]{
		\begin{minipage}[t]{0.45\linewidth}
			\centering
			\includegraphics[width=0.65in]{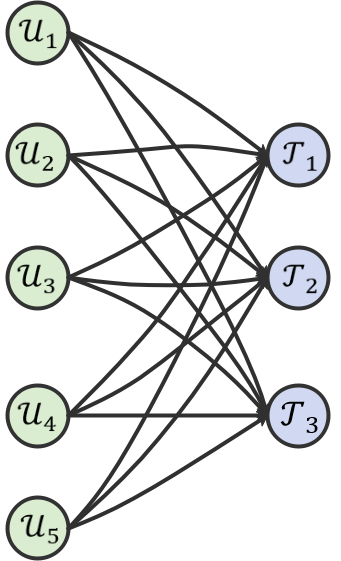}
			\label{fig3a}
		\end{minipage}%
	}%
	\subfigure[]{
		\begin{minipage}[t]{0.45\linewidth}
			\centering
			\includegraphics[width=1.4in]{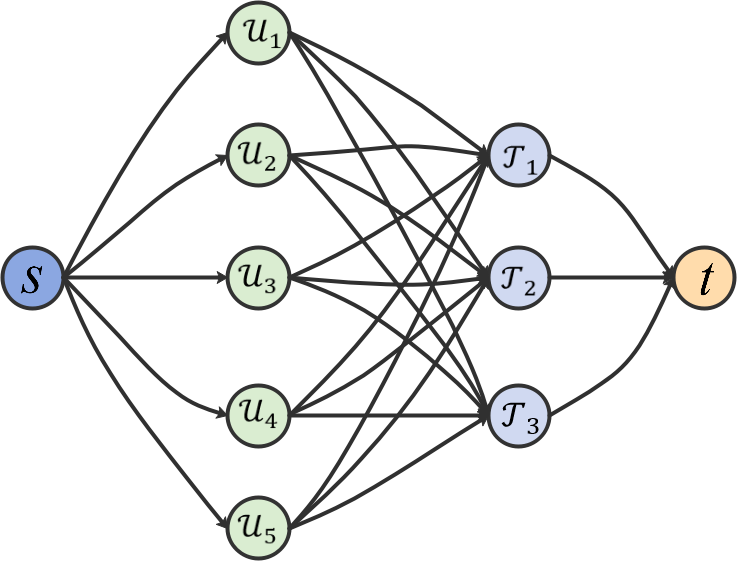}
			\label{fig3b}
		\end{minipage}%
	}%
	\caption{Graphs consisting of UAV vertices and target vertices.}
\end{figure}

We first build a graph $\mathcal G (\mathcal V, \mathcal A)$ (See Fig.~\ref{fig3a}) with the UAV vertices $\{\nu_{{\mathcal U}_{i}}\}_{i \in I_u}$, the target vertices $\{\nu_{{\mathcal T}_{j}}\}_{j \in I_\tau}$, and the edges $\{(\nu_{{\mathcal U}_{i}}, \nu_{{\mathcal T}_{j}}) \mid i\in I_u, j \in I_\tau \}$ connecting each UAV vertex and each target vertex. Then, graph $\mathcal G ({\mathcal V}_{st}, {\mathcal A}_{st})$ (See Fig.~\ref{fig3b}) can be obtained by adding a source vertex $\nu_s$, a sink vertex $\nu_t$, and new edges $\{(\nu_{s}, \nu_{{\mathcal U}_{i}})\}_{i \in I_u} $ and $\{(\nu_{{\mathcal T}_{j}},\nu_t)\}_{j \in I_\tau}$.
\begin{figure}[htbp]%
	\centering
	\subfigure[${\mathcal {\overline G}}_l({\mathcal V}_{st}, {\mathcal A}_{st})$]{
		\begin{minipage}[t]{0.33\linewidth}
			\centering
			\includegraphics[width=1.35in]{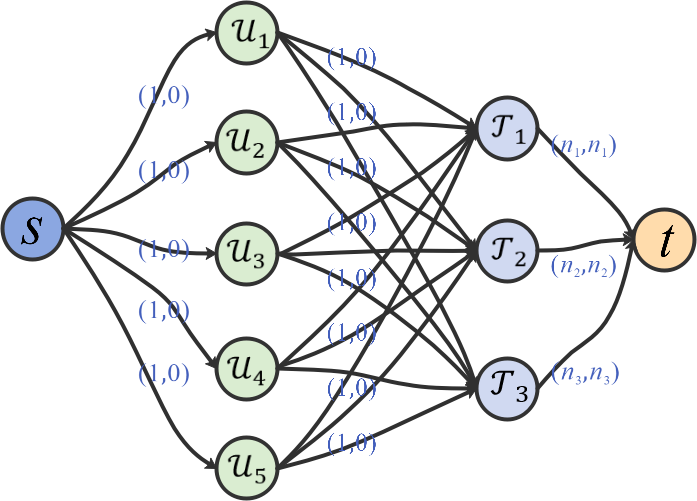}
			\label{fig4a}
		\end{minipage}%
	}%
	\subfigure[${\mathcal {\overline G}}_m({\mathcal V}_{st}, {\mathcal A}_{st})$]{
		\begin{minipage}[t]{0.34\linewidth}
			\centering
			\includegraphics[width=1.4in]{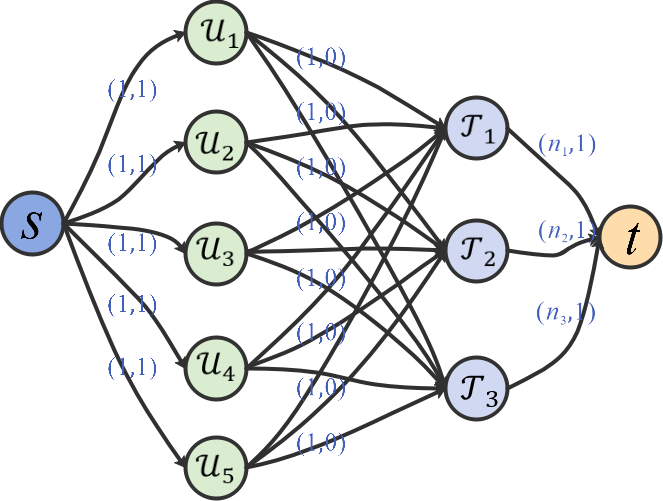}
			\label{fig4b}
		\end{minipage}%
	}%
	\subfigure[${\mathcal {\overline G}}_s({\mathcal V}_{st}, {\mathcal A}_{st})$]{
		\begin{minipage}[t]{0.33\linewidth}
			\centering
			\includegraphics[width=1.4in]{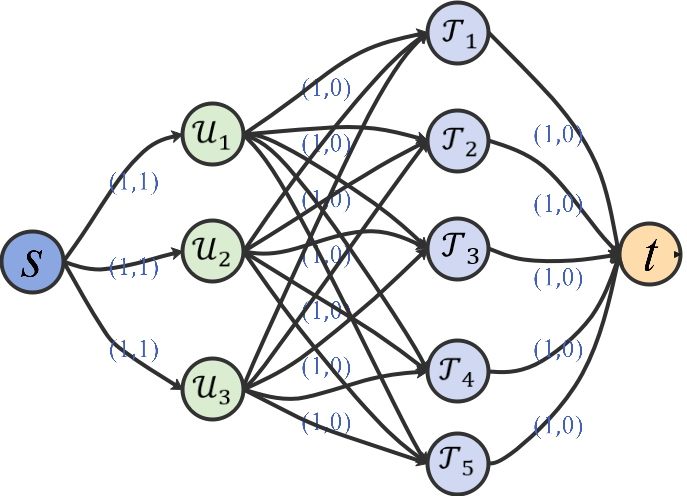}
			\label{fig4c}
		\end{minipage}%
	}%
	\caption{The network flow models with upper and lower flow bounds for multi-UAV task assignment according to the relationship between the number of UAVs and targets. (a)$N_u \geq  \sum \nolimits_{j=1}^{N_\tau} {n_j}$, (b)$N_\tau < N_u <  \sum \nolimits_{j=1}^{N_\tau} {n_j}$, and (c) $N_\tau \geq N_u$.}\label{fig4}
\end{figure}

Considering the relationship between the number of UAVs $N_u$, the number of targets $N_\tau$, and the maximum number of UAVs required for all targets $\sum \nolimits_{j=1}^{N_\tau} {n_j}$, graph $\mathcal G({\mathcal V}_{st}, {\mathcal A}_{st})$ can be further transformed into a capacity network ${\mathcal {\overline G}}({\mathcal V}_{st}, {\mathcal A}_{st})$ by adding some flow bounds (See Fig.~\ref{fig4}).
According to the optimization objective function (\ref{eq1}), our goal is to maximize the task rewards of the multi-UAV system. Then we set the cost of arc $(\nu_{{\mathcal U}_{i}},\nu_{{\mathcal T}_{j}})$ as $l(\nu_{{\mathcal U}_{i}},\nu_{{\mathcal T}_{j}}) = - R_{ij}$ and the cost of all other arcs as 0. The flow bounds and cost for each arc in the transformed network is shown in Table~\ref{tab_2}.  
\begin{table}[htbp]
	\begin{center}
		\begin{minipage}{1\textwidth}
			\caption{The flow bounds and cost for each arc in ${\mathcal {\overline G}}({\mathcal V}_{st}, {\mathcal A}_{st})$} 
			\footnotesize
			\renewcommand{\arraystretch}{1.4}
			\setlength{\tabcolsep}{1mm}
			\centering
			\begin{tabular}{cccccc} 
				\hline
				\multirow{2}*{\textbf{Numerical relationship}} &
				\multicolumn{3}{c}{\textbf{Flow bounds/}\boldmath{$ (upper,lower)$}} & \multicolumn{2}{c}{\textbf{Cost}}\\
				\cline{2-6}
				&\boldmath{$(\nu_{{\mathcal U}_{i}},\nu_{{\mathcal T}_{j}})$} & \boldmath{$(\nu_{s}, \nu_{{\mathcal U}_{i}})$} & \boldmath{$(\nu_{{\mathcal T}_{j}},\nu_t)$} & \boldmath{$(\nu_{{\mathcal U}_{i}},\nu_{{\mathcal T}_{j}})$} & \boldmath{$\mathcal A \backslash (\nu_{{\mathcal U}_{i}},\nu_{{\mathcal T}_{j}})$}\\
				
				\hline 
				$N_u \geq  \sum \nolimits_{j=1}^{N_\tau} {n_j}$ & (1,0) &(1,0) & ($n_j,n_j$)  &$-R_{ij}$  & 0 \\
				\hline
				$N_\tau < N_u <  \sum \nolimits_{j=1}^{N_\tau} {n_j}$ & (1,0) &(1,1) & ($n_j,1$) &$-R_{ij}$ & 0  \\
				\hline 
				$N_\tau \geq N_u$ & (1,0) &(1,1) & (1,0) &$-R_{ij}$ & 0 \\
				\hline
			\end{tabular}
			\label{tab_2} 
		\end{minipage}
	\end{center}
\end{table}
\begin{definition}[Network flow model for multi-UAV task assignment]\label{def_1}
	Given a graph $\mathcal G ({\mathcal V}_{st}, {\mathcal A}_{st})$ constructed by all UAV vertices and target vertices, set the upper flow bound of arc $(\nu_{{\mathcal U}_{i}}, \nu_{{\mathcal T}_{j}})$ as 1 and the lower as 0. According to the numerical relationship between $N_u$, $N_\tau$, and $\sum \nolimits_{j=1}^{N_\tau} {n_j}$, add the flow bounds and the cost to arc $(\nu_{s}, \nu_{{\mathcal U}_{i}})$ and arc $(\nu_{{\mathcal T}_{j}},\nu_t)$ with Table~\ref{tab_2}. The resulting capacity network ${\mathcal {\overline G}}({\mathcal V}_{st}, {\mathcal A}_{st})$ is a network flow model with upper and lower flow bounds for multi-UAV task assignment.
\end{definition}

For the transformed network, the feasible flow that satisfies the flow constraints must be the maximum flow. Moreover, when the network cost is minimized, the reward of the corresponding multi-UAV system is maximized. So far, the multi-UAV task assignment problem has been transformed into an MCMF problem.

\subsection{Task assignment minimum-cost maximum-flow algorithm}
The traditional MCMF algorithm is aimed at the case where there is no lower flow bound. Therefore, we further convert the network obtained in Section \ref{sec_4_2} and design an algorithm named task assignment minimum-cost maximum-flow (TAMM) for our multi-UAV task assignment problem. The detailed TAMM algorithm is given in Algorithm~\ref{algo_2}. In order to describe the algorithm more clearly, we first define the sum of the lower flow bounds of all arcs flowing into vertex $\nu_i$ and that flowing out of $\nu_i$ as $f_{in}^i$ and $f_{out}^i$.

\renewcommand{\algorithmicrequire}{\textbf{Input:}}
\renewcommand{\algorithmicensure}{\textbf{Output:}}
\begin{algorithm}[htbp!]
	\caption{Task assignment minimum-cost maximum-flow algorithm}\label{algo_2}
	\begin{algorithmic}[1]
		\Require Task reward matrix $R$ ($R$ is calculated in Section~\ref{sec5}), Network ${\mathcal {\overline G}}({\mathcal V}_{st}, {\mathcal A}_{st})$ (${\mathcal {\overline G}}$ is obtained by Definition~\ref{def_1})
		\Ensure The allocated task ${\mathcal T}_{j}$ for each UAV  
		\State  Add an arc $(\nu_t,\nu_s)$ with $upper_{ts}= + \infty$ and $lower_{ts}=0$, a source vertex $\nu_{s'}$, and a sink vertex $\nu_{t'}$. 
		\State The current edge set is ${\mathcal A}_{s't'} = {\mathcal A}_{st} \cup (\nu_t,\nu_s) $ and the current vertex set is ${\mathcal V}_{s't'} = {\mathcal V}_{st} \cup \{\nu_t',\nu_s'\}$;
		\For {$\forall \nu_i \in {\mathcal V}_{st}$}
		\If {$f_{in}^i \geq f_{out}^i$}
		\State For $\forall \nu_q \in {\mathcal V}_{s't'}$ with {$lower_{qi} \textgreater 0$}, let $upper_{qi}= upper_{qi} - lower_{qi}$ and $lower_{zi} = 0$;
		\State Add an arc $(\nu_{s'},\nu_i)$ with $upper_{s'i}=f_{in}^i - f_{out}^i$ and $lower_{s'i}=0$. Then ${\mathcal A}_{s't'} = {\mathcal A}_{s't'}\cup (\nu_{s'},\nu_i)$;
		\Else
		\State For $\forall \nu_p \in {\mathcal V}_{s't'}$ with {$lower_{ip} \textgreater 0$}, let $upper_{ip}= upper_{ip} - lower_{ip}$ and $lower_{ip} = 0$;
		\State Add an arc $(\nu_i,\nu_{t'})$ with $upper_{it'}=f_{in}^i - f_{out}^i$ and $lower_{it'}=0$. Then, ${\mathcal A}_{s't'} = {\mathcal A}_{s't'}\cup (\nu_i,\nu_{t'})$;
		\EndIf
		\EndFor
		\State For { $\forall (\nu_i, \nu_j) \in {\mathcal A}_{s't'}$}, delete $(\nu_i, \nu_j)$ with {$upper_{ij}=0$ and $lower_{ij} = 0$};
		\For { $\forall \nu_i \in {\mathcal V}_{s't'}$}
		\If {the in-degree and out-degree of $\nu_i$ are both 1}
		\State Find $(\nu_j, \nu_i), (\nu_i, \nu_k) \in {\mathcal A}_{s't'}$, then delete $\nu_i$, $(\nu_j, \nu_i)$, and $(\nu_i, \nu_k)$; 
		\State Add the arc $(\nu_j,\nu_k)$ with $c(\nu_j,\nu_k) = \max \{ c(\nu_j,\nu_i), c(\nu_i,\nu_k)\}$;
		\EndIf
		\EndFor
		\State 	Denote the current network as ${\mathcal {\hat G}}({\mathcal V}_{s't'}, {\mathcal { A}}_{s't'})$;
		\State Run a minimum-cost maximum-flow method to find the MCMF $f_{mcmf}$ of ${\mathcal {\hat G}}({\mathcal V}_{s't'}, {\mathcal { A}}_{s't'})$; 
		\For {$\forall i \in I_u$}
		\If {$\exists j \in I_\tau, {f_{{{\mathcal U}_{i}}{{\mathcal T}_{j}}} = 1}$ in $f_{mcmf}$ } 
		\State ${\mathcal U}_{i}$ is assigned to track target ${\mathcal T}_{j}$;
		\EndIf
		\If {$\sum_{j \in I_\tau} {f_{{{\mathcal U}_{i}}{{\mathcal T}_{j}}} = 0}$ in $f_{mcmf}$}
		\State ${\mathcal U}_{i}$ is assigned to cover the area. Let ${\mathcal T}_{0}$ denote the area coverage task.
		\EndIf
		\EndFor
	\end{algorithmic}
\end{algorithm}

\begin{figure}[htbp]%
 	\centering
 	\includegraphics[width=1\textwidth]{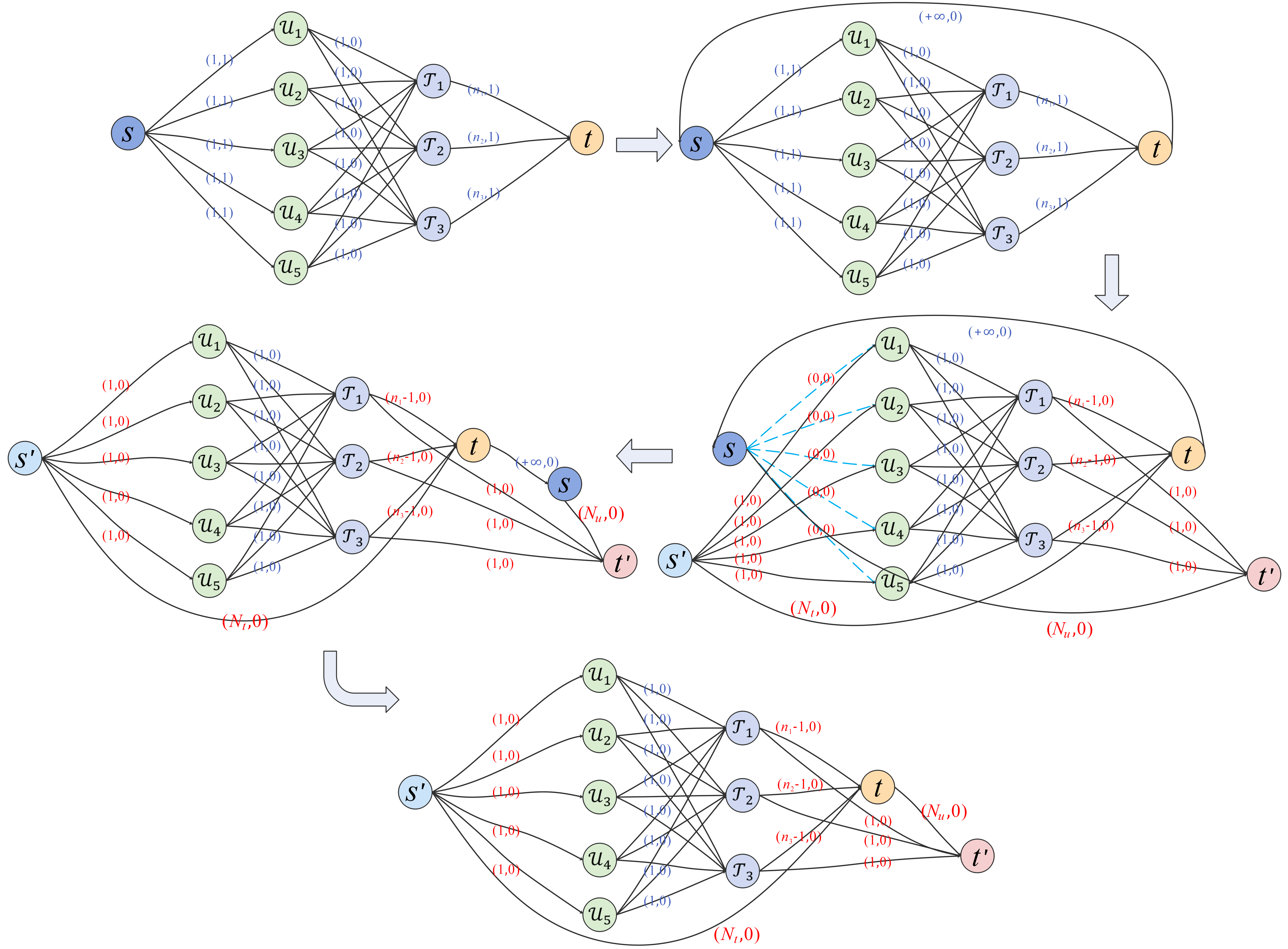}
 	\caption{The detailed conversion process to get network ${\mathcal {\hat G}}({\mathcal V}_{s't'}, {\mathcal { A}}_{s't'})$ without lower flow bounds.}\label{fig5}
 \end{figure}

The basic idea of TAMM is to convert the network with lower flow bounds ${\mathcal {\overline G}}({\mathcal V}_{st}, {\mathcal A}_{st})$ into the network ${\mathcal {\hat G}}({\mathcal V}_{s't'}, {\mathcal { A}}_{s't'})$ without lower flow bounds by adding new source vertex, sink vertex, and related edges. The detailed conversion process is shown in Fig.~\ref{fig5}.
Then the minimum-cost maximum-flow method can be used to find the MCMF of the converted network.  The task assignment results can be obtained by mapping the obtained flow to the matching relationship between UAVs and targets. 
\begin{remark}
	Our task assignment algorithm entails a complexity  ${\mathcal O}(\vert {\mathcal V}_{s't'} \vert \vert {\mathcal { A}}_{s't'} \vert)$. Since $\vert {\mathcal V}_{s't'} \vert = N_u + N_\tau +3$ and $\vert {\mathcal { A}}_{s't'} \vert = N_u + N_u N_\tau + 2(N_\tau + 1)$ in network ${\mathcal {\hat G}}({\mathcal V}_{s't'}, {\mathcal { A}}_{s't'})$, the complexity of task assignment is $\mathcal O ({N_u^2}N_\tau + N_u{N_\tau^2})$. Therefore, our TAMM algorithm has the polynomial time complexity and can quickly solve the multi-UAV task assignment problem. 
\end{remark}
 Since the MCMF of ${\mathcal {\overline G}}({\mathcal V}_{st}, {\mathcal A}_{st})$ corresponds to the allocation plan that maximizes the task reward.
To illustrate the MCMF of the converted  network ${\mathcal {\hat G}}({\mathcal V}_{s't'}, {\mathcal { A}}_{s't'})$ also corresponds to the allocation plan with the maximum task reward, we introduce the following theorem.
\begin{theorem}\label{thm0}
	Given the network ${\mathcal {\overline G}}({\mathcal V}_{st}, {\mathcal A}_{st})$ with the upper and lower bounds, construct the network following the steps in Algorithm~\ref{algo_2}, then the minimum-cost maximum-flow of the original network ${\mathcal {\overline G}}({\mathcal V}_{st}, {\mathcal A}_{st})$ is also that of the converted network ${\mathcal {\hat G}}({\mathcal V}_{s't'}, {\mathcal { A}}_{s't'})$. 
\end{theorem}

The proof of Theorem~\ref{thm0} is based on the basic concepts of network flow mentioned in Section~\ref{sec4_1}.

\begin{proof}
	In the transformation process, the original network is first converted into the no-source and no-sink network ${\mathcal {\overline G}}({\mathcal V}_{st}, {\mathcal {\overline A}}_{st})$ by introducing an arc $(\nu_t,\nu_s)$ with the capacity constraint $+ \infty$. The minimum-cost maximum-flow of the original network is also that of ${\mathcal {\overline G}}({\mathcal V}_{st}, {\mathcal {\overline A}}_{st})$.
	After that, each step of the transformation from network ${\mathcal {\overline G}}({\mathcal V}_{st}, {\mathcal {\overline A}}_{st})$ to network ${\mathcal {\hat G}}({\mathcal V}_{s't'}, {\mathcal { A}}_{s't'})$ follows the conservation constraints (\ref{eq17}).
	
	Then, if a feasible flow $f$ makes any arcs $(\nu_{s'},\nu_i)$ or $(\nu_i,\nu_{t'})$ in ${\mathcal {\hat G}}({\mathcal V}_{s't'}, {\mathcal { A}}_{s't'})$  reach maximum capacity, this flow must be a feasible flow of network ${\mathcal {\overline G}}({\mathcal V}_{st}, {\mathcal {\overline A}}_{st})$. On the contrary, any feasible flow in network  ${\mathcal {\overline G}}({\mathcal V}_{st}, {\mathcal {\overline A}}_{st})$ corresponds to the flow whose arc $(\nu_{s'},\nu_i)$ or $(\nu_i,\nu_{t'})$ achieves maximum capacity in network ${\mathcal {\hat G}}({\mathcal V}_{s't'}, {\mathcal { A}}_{s't'})$.
	
	The flow reaching maximum capacity from the source $\nu_{s'}$ to the sink $\nu_{t'}$ of network ${\mathcal {\hat G}}({\mathcal V}_{s't'}, {\mathcal { A}}_{s't'})$ must be the maximum flow of the network. Therefore, finding a feasible flow of network ${\mathcal {\overline G}}({\mathcal V}_{st}, {\mathcal {\overline A}}_{st})$ is equivalent to finding the maximum flow from $\nu_{s'}$ to $\nu_{t'}$ of network ${\mathcal {\hat G}}({\mathcal V}_{s't'}, {\mathcal { A}}_{s't'})$.
	
	According to the flow constraints of $(\nu_s,\nu_i)$ and $(\nu_i,\nu_t)$ in network ${\mathcal {\overline G}}({\mathcal V}_{st}, {\mathcal {\overline A}}_{st})$, the feasible flow of ${\mathcal {\overline G}}({\mathcal V}_{st}, {\mathcal {\overline A}}_{st})$ must be its maximum flow. Then the maximum flow of the network ${\mathcal {\hat G}}({\mathcal V}_{s't'}, {\mathcal { A}}_{s't'})$ is also the maximum flow of ${\mathcal {\overline G}}({\mathcal V}_{st}, {\mathcal {\overline A}}_{st})$. Since the cost of the newly added arc is 0, the minimum-cost maximum-flow $f_m$ of the converted network ${\mathcal {\hat G}}({\mathcal V}_{s't'}, {\mathcal { A}}_{s't'})$ is that of ${\mathcal {\overline G}}({\mathcal V}_{st}, {\mathcal {\overline A}}_{st})$. Furthermore, $f_m$ is also the minimum cost maximum-flow of the original network ${\mathcal {\overline G}}({\mathcal V}_{st}, {\mathcal A}_{st})$.
\end{proof}

\section{Multi-UAV behavior decision-making}\label{sec5}
In our modular framework, the cooperative behavior decision-making process includes tracking behavior decision-making module and coverage behavior decision-making module. 
The main idea of cooperative behavior decision-making is to first get the tracking action and the corresponding rewards through the tracking decision-making for each UAV and then assign tasks to UAVs according to the reward matrix. If a UAV is assigned one target to track, its action is the optimal tracking action calcaulated by tracking decision-making module. While if a UAV is assigned no target to track, it then plans its action through coverage decision-making.

\subsection{Tracking behavior decision-making}
The objective of tracking decision-making is to plan the actions of UAVs according to the predicted target states to obtain more accurate measurements of targets.

We use the determinant of the Fisher information matrix (FIM) as the reward value. \cite{bib34} explained the reasons for using FIM as the reward function:  FIM processes the original measurement data from the sensor and is derived from the measurement model in the natural sensor polar coordinate system directly, reflecting the volume of information of the measurement data. 

The FIM of UAV ${\mathcal U}_i$ about target ${\mathcal T}_j$ at time $k$ is defined as $G_{ij}(k)$, and $det(G_{ij}(k))$ denotes the determinant of the FIM.
When considering the information accumulated over $H$ time step, the reward at time step $k$ can be approximated by the sum of the determinants of the FIM as follows:
\begin{equation}
	R_{ij} = \sum_{l=1}^{H} { [det(G_{ij}({k+l}))]}
\end{equation}

The tracking decision-making algorithm based on the rolling horizon method finds the optimal action sequence that maximizes the cumulative Fisher information volume.
\begin{equation}
	 \max \limits_{({\hat \pi}_i(k),\cdots,{\hat \pi}_i({k+H-1}))} {\sum_{l=1}^{H} { [det(G_{ij}({k+l}))]}}
\end{equation}

 At time step $k$, each UAV predicts the target positions of the future $H$ steps according to the state $\overline{s}_{i,j}(k)$ updated in the distributed fusion phase. Then the optimal planning method is used to find the optimal action sequence ${(\pi_i^{opt}(k),\cdots,\pi_i^{opt}(k+H-1))}$ that maximizes the objective function. The planned action of ${\mathcal U}_i$ is the first item of the optimal action sequence, which is:
 
 \begin{equation}
 	\pi_i(k) = \pi_i^{opt}(k)
 \end{equation}
We also call the planned action the optimal tracking action. The optimal tracking action of UAVs at each step can be obtained by repeating the above process.  
 
\subsection{Coverage behavior decision-making}
The purpose of the area coverage task is to continuously search the targets and reduce the uncertainty of the dynamic environment. Inspired by solitary organisms, the distributed anti-flocking algorithm \cite{bib20} is well suited for area coverage tasks and is summarized as follows:
 1) Collision avoidance: avoid surrounding obstacles and neighbors. 2) Decentralization: move away from neighboring centers. 3) Selfishness: maximize own gains.
 
 Our previous work \cite{bib18} designed a distributed multi-UAV persistent coverage algorithm based on the anti-flocking method, of which several area coverage information maps are designed. 
 The algorithm uses the collision avoidance and decentering rules in the anti-flocking method to design a heading matching map (HMM), making the UAVs disperse from each other to increase the instantaneous coverage area and avoid collisions. 
 The EVTM, the compressed EVTM, and the HMM are integrated to calculate the detection rewards of UAVs and guide the UAV to the area where the rewards are maximized. We briefly describe the algorithm below.

First, calculate the desired separation heading $\eta_{i}^d$ of each UAV through the rules of collision avoidance and decentering. 

\begin{equation}
	\begin{split}
		\eta_{i}^{d} = &k_{o} \left[ \sum\limits_{o=1}^{M_i}{\mathcal S(\left\| {{p}_{i}}-p_{i}^{o} \right\|,{d_{i}}^o)\frac{{{p}_{i}}-p_{i}^{o}}{\left\| {{p}_{i}}-p_{i}^{o} \right\|}} \right] + \\ & k_{c} \left[ \mathcal S(\left\| {{p}_{i}}-p_{i}^{c} \right\|,{d_{c}})\frac{{{p}_{i}}-p_{i}^{c}}{\left\| {{p}_{i}}-p_{i}^{c} \right\|} \right]
	\end{split}
	\label{eq23}
\end{equation}
  $\mathcal S(\cdot)$ is a non-negative repulsive potential function. $d_{i}^{o}$ represents the safety distance of ${\mathcal U}_i$, $d_c$ is the distance threshold that the decentering term works, and $M_i$ is the total number of neighboring obstacles and UAVs of ${\mathcal U}_i$. Besides, $p_{i}^{c}$ denotes the position of the centroid of the neighbor UAVs.

Let $\eta_{i}^{m,n}$  denote the heading of ${\mathcal U}_i$ to the grid $g(m,n)$, that is:

\begin{equation}
	\eta_{i}^{m,n} = \frac{{{p}_{m,n}}-p_{i}}{\left\| {{p}_{m,n}}-p_{i} \right\|}
\end{equation}
where $p_{m,n}$ represents the position of grid $g(m,n)$.

The heading matching value between $\eta_{i}^{m,n}$ and the desired heading $\eta_{i}^d$ is:

\begin{equation}
	{A}_{i}^{m,n}\!=\!
	\left\{
	\begin{aligned}
		& e^{-k_{a} \left(\eta_{i}^{m,n} - \eta_{i}^{d} \right)^2 } & \vert \Delta \eta_{i}^{m,n} \vert < {\omega}_{max}\Delta T \\ 
		& 0 & otherwise
	\end{aligned} 
	\right.
\end{equation}
where $\Delta\eta_{i}^{m,n}$  is the deviation value between $\eta_{i}^{m,n}$ and $\eta_{i}$. 

Then the coverage reward is calculated from EVTM ${T}_i$. We define the coverage reward that ${\mathcal U}_i$ obtains from grid $g(m,n)$ by the visiting requirement $\lambda_{i}^{m,n}$ as follows: 

\begin{equation}
	f_{i}^{m,n} = \lambda_{i}^{m,n}
	\label{eq26}
\end{equation}

Suppose ${\mathcal U}_i$ is located at $g(m,n)$ at the next step and $\phi_{s_{i} (m,n)}$ denotes the detection area of ${\mathcal U}_i$, then the predicted coverage reward is:

\begin{small}
	\begin{equation}
		{ F}_{i}^{m,n} = \sum_{p_{m',n'} \subset \phi_{s_{i} (m,n)} }\lambda_{i}^{m',n'} 
	\end{equation}
\end{small}

The global searching reward map  of ${\mathcal U}_i$ is $Q_i \in {\mathbb{R}}^{3 \times 3}$, which calculated from the compressed EVTM ${T}_{i}^{g}$ as follows:

\begin{equation}
	{ Q}_{i}=(t_{k} {\bf 1}_{3}-{T}_{i}^{g})/T_{c}
\end{equation}

Furthermore, we need to extract the heading matching value and the coverage reward of the $3 \times 3$ grids around the location of the UAV to get the local heading matching map ${A}_i^l$ and local coverage reward map ${F}_i^l$. The overall coverage reward map (OCRM) $J_i \in {\mathbb{R}}^{3 \times 3}$ consists of the weighted sum of $F_{i}^l$, $Q_{i}$, and $A_{i}^l$, which is:

\begin{equation}
	J_{i}(\kappa,\iota)=w_{f}F_{i}^{l}(\kappa,\iota) + w_{q}Q_{i}(\kappa,\iota) + w_{a}A_{i}^{l}(\kappa,\iota) \quad 
	\label{eq29}
\end{equation} 
where $w_{f}$, $w_{q}$ and $w_{a}$ are the weights of the corresponding term.

Based on the selfishness rule of the anti-flocking method, we directly select the grid with maximum reward in the OCRM as the target grid $g(m_*,n_*)$.
$(\kappa',\iota')=\mathop{\arg \max}\limits_{(\kappa,v)} J_{i}(\kappa,\iota) $ denotes the grid with the maximum reward in map $J_i$. When ${\mathcal U}_i$ is located at $g(m_i,n_i)$, then $m_*=\kappa'+m_i-2$ and $n_*=\iota'+n_i-2$.

Finally, let $g(m_*,n_*)$ be the desired gird of ${\mathcal U}_i$. The control input $\pi_i = (\Delta v_i,\omega_{i})$ of $U_i$ is: 

\begin{small}
	\begin{equation}
		\left\{
		\begin{aligned}
			& \Delta v_{i}=\left[\|p_{m_*,n_*}-p_{i}\|/ \Delta T - v_{i}
			\right]_{-dv_{max}}^{dv_{max}} \\
			& \omega_{i}=\left[ (\eta_{i}^{*}-\eta_{i})/ \Delta T
			\right]_{-\omega_{max}}^{\omega_{max}}\\
		\end{aligned}
		\right.
		\label{eq31}
	\end{equation}
\end{small}
$p_{m_*,n_*}$ is the position of grid $g(m_*,n_*)$. And $\eta_{i}^{*}$ is the desired heading when ${\mathcal U}_i$ flies to $g(m_*,n_*)$ from its current position, which is:

\begin{equation}
	\eta_{i}^{*}=atan2(y_i^{*}-y_{i},x_i^{*}-x_{i})
	\label{eq32}
\end{equation}

\section{System design of distributed multi-UAV cooperative area coverage and target tracking}\label{sec6}
\begin{figure}[htbp]%
	\centering
	\includegraphics[width=0.7\textwidth]{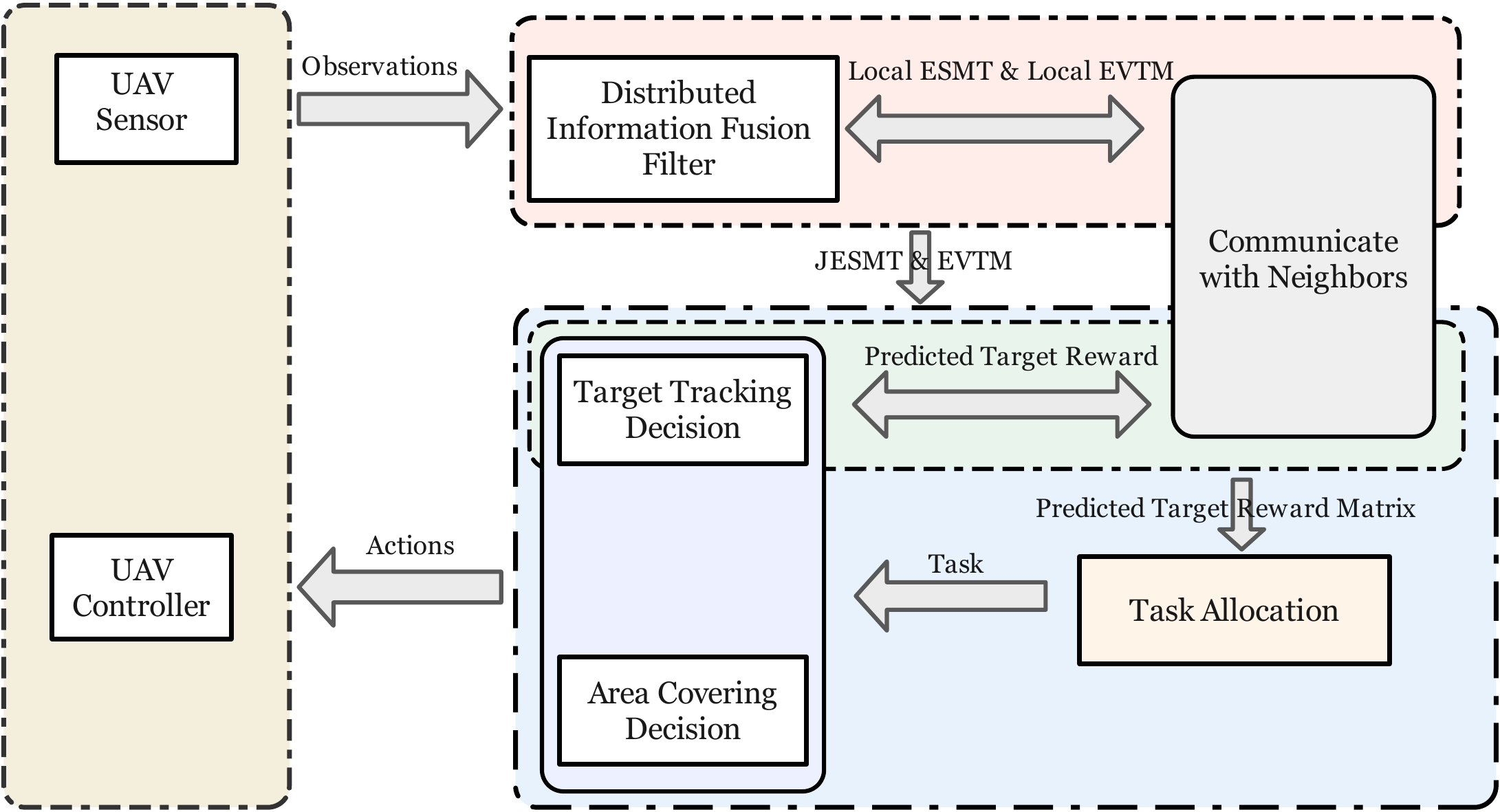}
	\caption{The distributed multi-UAV cooperative area coverage and target tracking framework.}\label{fig6}
\end{figure}

Integrating the information fusion strategy, the task assignment method, and the behavior decision algorithm designed in Sections \ref{sec3}-\ref{sec5}, this section designs the distributed multi-UAV cooperative area coverage and target tracking algorithm. The framework is shown in Fig.~\ref{fig6}.

As shown in the information fusion process, we use the maximum consensus protocol to update the joint estimation state of multi-targets and the equivalent visiting time map, which reach the consensus in the connected communication network after limited iterations. Then each UAV uses the Fisher information reward value to get the optimal tracking action sequence and exchanges the Fisher information reward with its neighbors to get the reward matrix.
The task assignment method based on minimum-cost maximum-flow assigns the task to each UAV that maximizes the task reward. If ${\mathcal U}_i$ is assigned a target to track, the optimal tracking action of ${\mathcal U}_i$ is the first item of the optimal tracking action sequence. While for those UAVs assigned the coverage task, we use the coverage behavior decision-making algorithm based on the anti-flocking method to plan its action.

In particular, the task assignment algorithm in Section \ref{sec4} needs to know the global reward matrix. While each UAV only has its reward about targets initially. Therefore, each UAV exchanges the Fisher information reward with its neighbors. Through limited updates with max-consensus protocol, the reward matrix of all UAVs in the connected network achieves consistency. 
As with the information fusion algorithm in Section \ref{sec3}, we set the iteration number for the $i$th UAV as ${\mathcal D}_i$ according to the termination condition of the maximum consistency algorithm given in \cite{bib34}. Therefore, the distributed information interaction enables the proposed task allocation algorithm to be adapted to the overall distributed framework.

Algorithm \ref{algo_3} below summarizes the distributed approach for local detection, fusion, assignment, and decision-making of ${\mathcal U}_i$.

\begin{algorithm}[htbp!]
	\caption{Local detection, fusion, assignment, and decision-making for the $i$th UAV at time $k$}\label{algo_3}
	\begin{algorithmic}[1]
		\Require ${T}_{i}(k-1), \{ \{ \overline{s}_{i,j}(k-1), \overline{P}_{i,j}(k-1) \} \mid j \in I_\tau \}   $
		\Ensure Action $\pi_i(k)$
		\State Run Algorithm \ref{algo_1} to get $ {T}_{i}(k), {T}_{i}^l(k),{T}_{i}^g(k), \{ \{ \overline{s}_{i,j}(k), \overline{P}_{i,j}(k) \} \mid j \in I_\tau \}  $ after information fusion;
		\For{$j = 1$ to $N_\tau$} 
		\State  $R_{ij}^{opt}(k) \gets 	 \max \limits_{({\hat \pi}_i({k}),\cdots,{\hat \pi}_i({k+H-1}))} {\sum_{l=1}^{H} { [det(G_{ij}(k+l))]}} $;
		\State $\pi_{ij}^{opt}(k) \gets \mathop{\arg\max} \limits_{({\hat \pi}_i(k),\cdots,{\hat \pi}_i({k+H-1}))} {\sum_{l=1}^{H} { [det(G_{ij}({k+l}))]}} $;
		\EndFor
		\State $R_i^{opt}(k) := [R_{i1}^{opt}(k), R_{i2}^{opt}(k), \cdots,  R_{i{N_\tau}}^{opt}(k)]$;
		\State $ R_{i}^{all}(0) := \{R_{i}^{opt}(k)\}$;
		\For{$\iota = 1$ to ${\mathcal D}_i$}
		\State Send $R_{i}^{all}(\iota-1)$;
		\State Receive $R_{q}^{all}(\iota-1), q \in \mathscr{V}_i^n, q \neq i$;
		\State $ R_{i}^{all}(\iota)  \gets \bigcup\limits_{ q \in \mathscr{V}_i^n}  R_{q}^{all}(\iota-1)$;
		\EndFor
		\State Get reward matrix $R(k)$ by the elements in $R_{i}^{all}({\mathcal D}_i)$;
		\State The assigned task got by Algorithm \ref{algo_2} is ${\mathcal T}_{j_*} \gets TAMM(R(k))$;
		\If{$j_* = 0$}
		\State Calculate the overall coverage reward map $J_i$ by (\ref{eq29});
		\State $(\kappa',\iota') \gets \mathop{argmax}\limits_{(\kappa,\iota)} J_{i}(\kappa,\iota) $;
		\State The target grid $g(m_*,n_*)$: $ m_*=\kappa'+m_i-2, n_*=\iota'+n_i-2 $;
		\State Get $\pi_i(k)$ by (\ref{eq31}) and (\ref{eq32});
		\Else
		\State  $\pi_i(k) \gets \pi_{i{j_*}}^{opt}(k)$.
		\EndIf
	\end{algorithmic}
\end{algorithm}

In the tracking behavior decision phase, each UAV calculates its rewards for all tasks. The complexity of a single step of computing is $\mathcal O (N_\tau \vert \Pi \vert)$, where $\vert \Pi \vert$ is  the cardinality of the action set. Compared with the computational complexity $\mathcal O (N_{u}N_{\tau}\vert \Pi \vert)$ of the centralized decision-making, our approach has lower computational complexity. Apart from the decision-making phase, the task assignment phase entails a complexity $\mathcal O ({N_u^2}N_\tau + N_u{N_\tau^2})$.
The information fusion phase also has low computational complexity. Therefore, our distributed hierarchical modular algorithm can be well applied to the area coverage and target tracking tasks of a large area with multiple UAVs.

\section{Simulation results}\label{sec7}
The proposed Algorithm \ref{algo_3} for area coverage and target tracking has been analyzed in the simulated environment. The area surveillance mission aims to maintain continuous coverage of the area and keep track of the detected targets. We evaluate the performance of our method through performance indicators such as area uncovered time and target observation coverage rate in a series of simulations.
The setup for the different simulations and their corresponding results are detailed in the following subsections.

\subsection{Simulation setup}\label{sec7_1}
Suppose the UAV flies at a fixed altitude of $90-110m$, and the communication radius $R_c$ between the UAV is set through specific experiments.
We set the task area to be $2.5km*2.5km$ which is rasterized into grids with both length and width of $20m$. $4$ ground moving targets are initially randomly distributed in this area, and their speed and heading can change with time. The sensor parameters are the same as those in \cite{bib34}. 
\begin{figure}[htbp!]%
	\centering
	\includegraphics[width=0.45\textwidth]{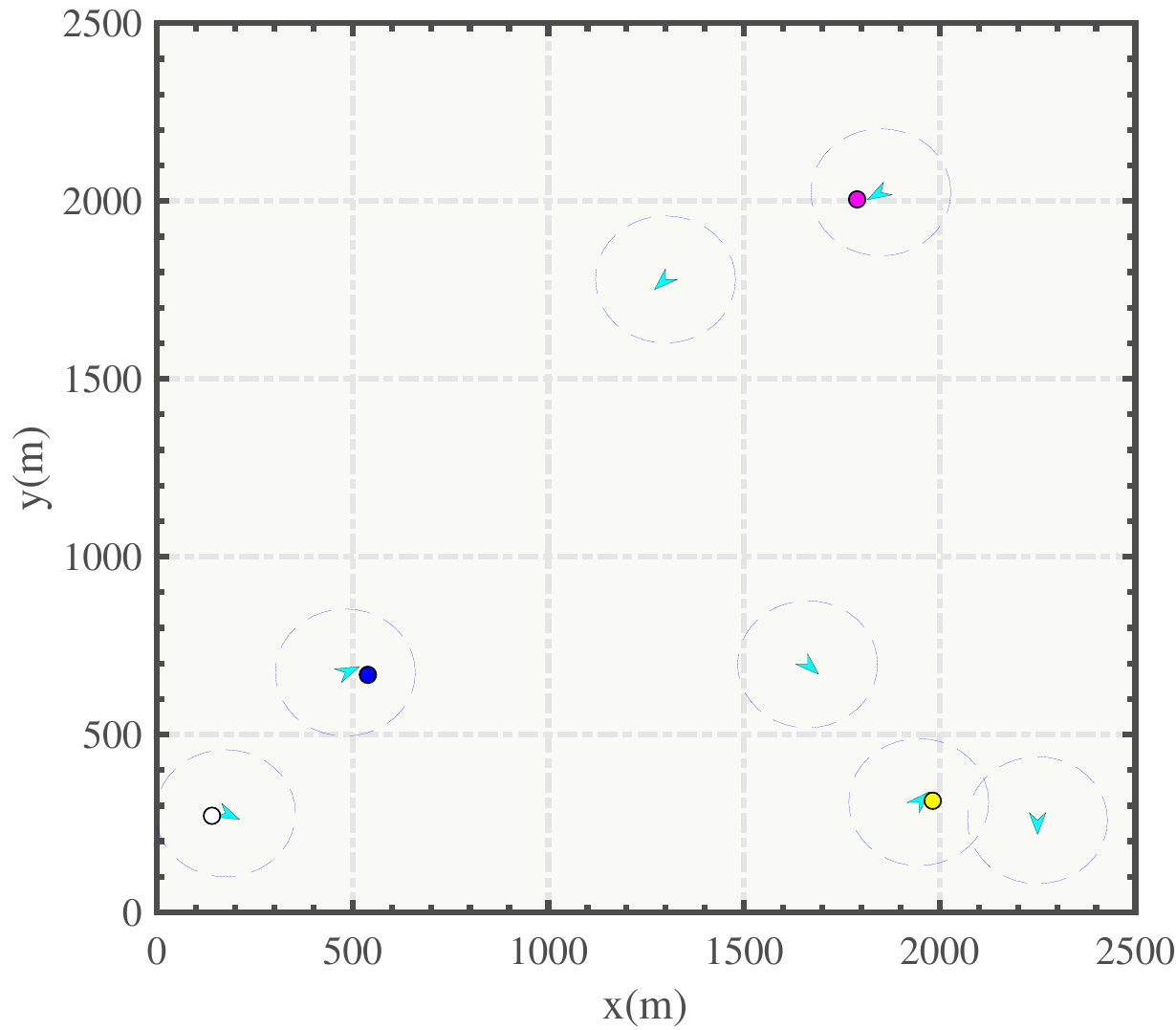}
	\caption{The simulation scenario with seven UAVs and four targets.}\label{fig7}
\end{figure}

The numerical simulation is implemented on MATLAB R2020b, and Fig.~\ref{fig7} is one example of the simulation scenarios. And we also perform the hardware-in-the-loop simulations, which are described in Section~\ref{sec7_3}. Assuming that each target can be tracked by a maximum of 2 UAVs simultaneously, set the parameters related to grid visiting requirements as $w_{p1} = 1,w_{p2} = 1.2,w_{p3}=0.8$ and the related parameters of sensor detection performance as  $\alpha = 1.1,\beta = 0.8$. Considering the randomness of the experiment, including the initial positions of the target and the UAV, and the random movement of the target, we performed 50 repetitions of each set of experiments with 1500 simulation steps to get average results.

\subsection{Simulation analysis}\label{sec7_2}

\subsubsection{The effect of different parameters}\label{sec7_21}
In this subsection, we take the simulations with different communication ranges and UAV numbers to study the effect of these parameters on the performance of our distributed area surveillance method.

\begin{figure}[htbp]%
	\centering
	\includegraphics[width=0.95\textwidth]{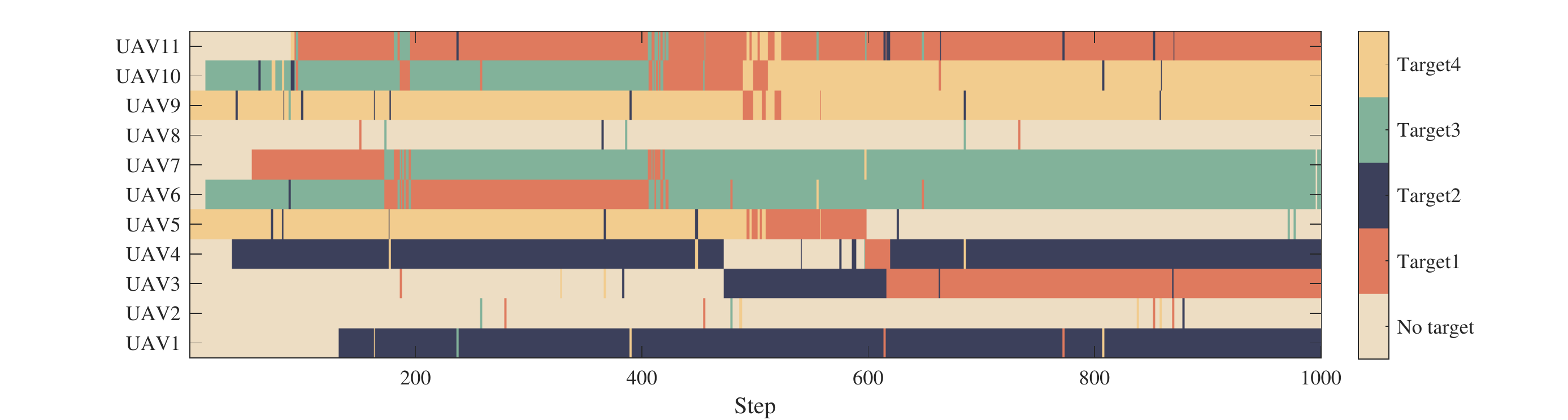}
	\caption{The allocation results when the number of UAVs is 11.}\label{fig9}
\end{figure}

First, we set the UAV numbers as $N_u\in[3,13]$ with a fixed communication radius of $400m$. 
Since the target is tracked by at most two UAVs, ideally, when $N_u < 4$,  there have some targets that are not assigned to the UAV. When $4\leq N_u\leq 8$, each target can be assigned to at least one UAV to track. When $N_u>8$, each target can be assigned to two UAVs to track. Fig.~\ref{fig9} is an example of the allocation result when the number of UAVs is 11. Due to the random movement of the target, the detected targets can be lost and require the multi-UAV system to search the targets again.

\begin{figure}[htbp]
	\centering
	\begin{minipage}[t]{0.48\textwidth}
		\centering
		\includegraphics[width=1.85in,height=1.5in]{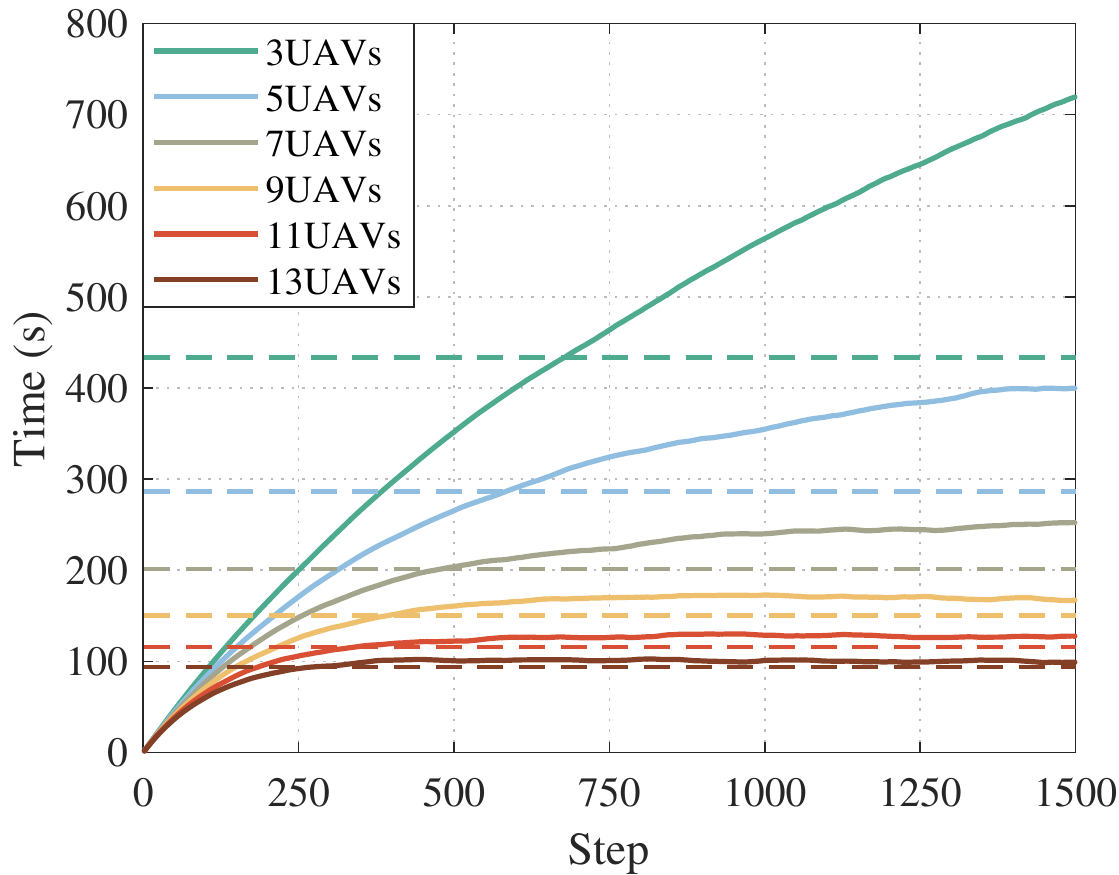} 
		\caption{The average uncovered time of all grids at each time step under different UAV numbers.}\label{fig10}
	\end{minipage}
	\begin{minipage}[t]{0.48\textwidth}
		\centering
		\includegraphics[width=1.85in,height=1.5in]{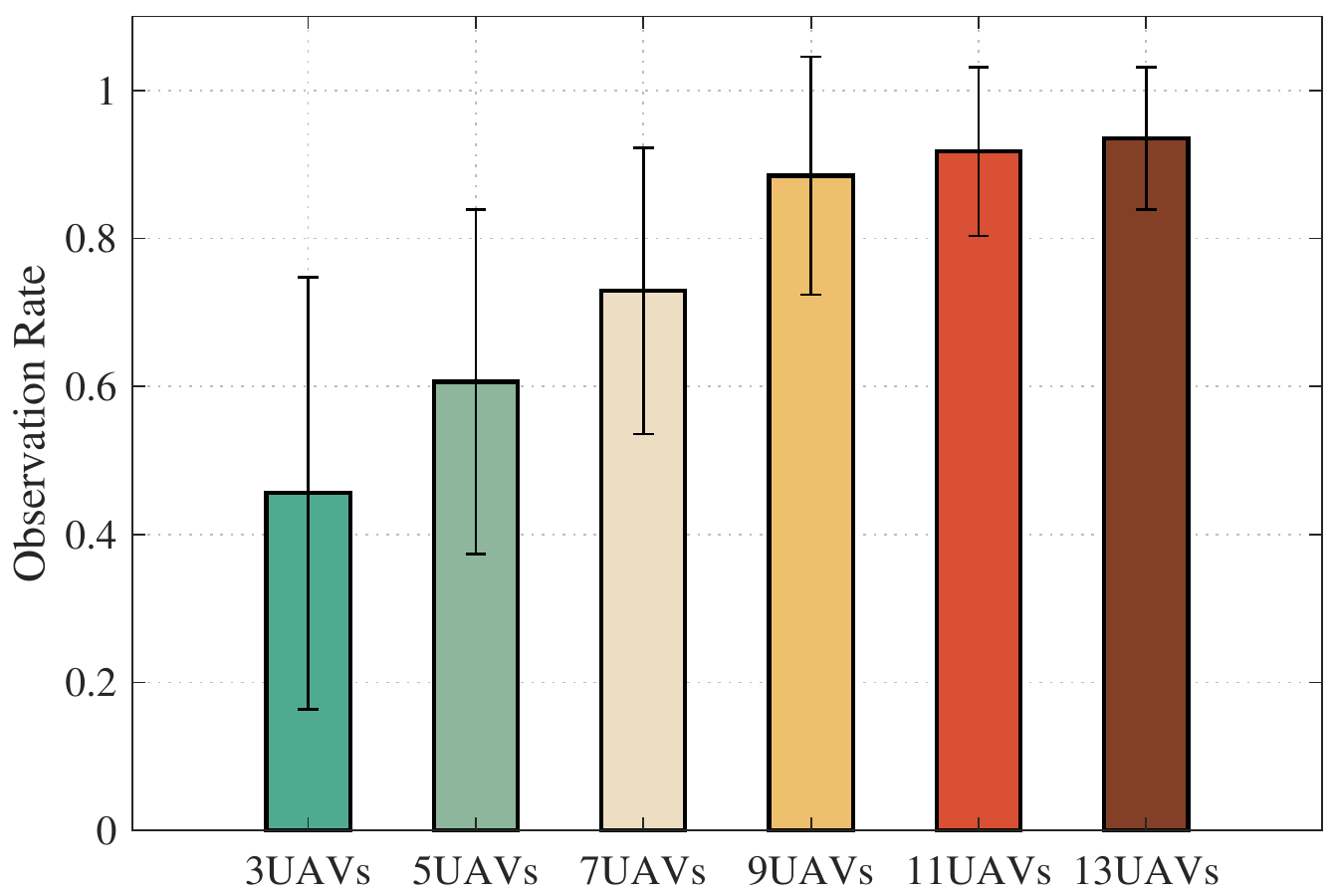} 
		\caption{The average observation coverage rate of all targets under different UAV numbers.}\label{fig12}
	\end{minipage}
\end{figure}

We use the instantaneous average uncovered time $t_{IMT}(k)$ to present the coverage performance of our algorithm, as shown in Fig.~\ref{fig10}. $t_{IMT}(k)$ indicates the average uncovered time of all grids at time step $k$. We find that the higher the number of UAVs, the shorter the average uncovered time, indicating a shorter visit interval.



\begin{figure}[htbp]%
	\centering
	\subfigure[]{
		\begin{minipage}[t]{0.48\linewidth}
			\centering
			\includegraphics[width=1.8in,height=1.5in]{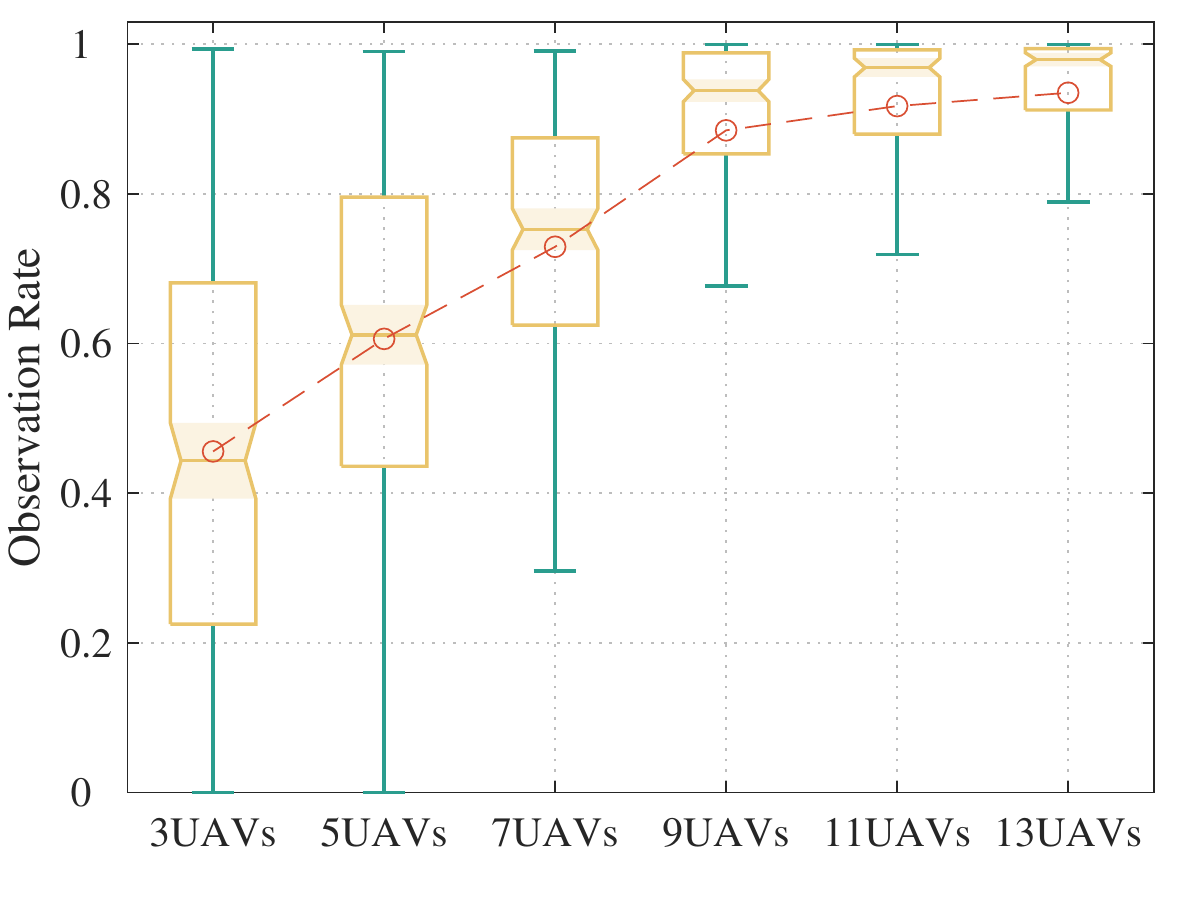} 
		\end{minipage}%
	}%
	\subfigure[]{
		\begin{minipage}[t]{0.48\linewidth}
			\centering
			\includegraphics[width=1.85in,height=1.5in]{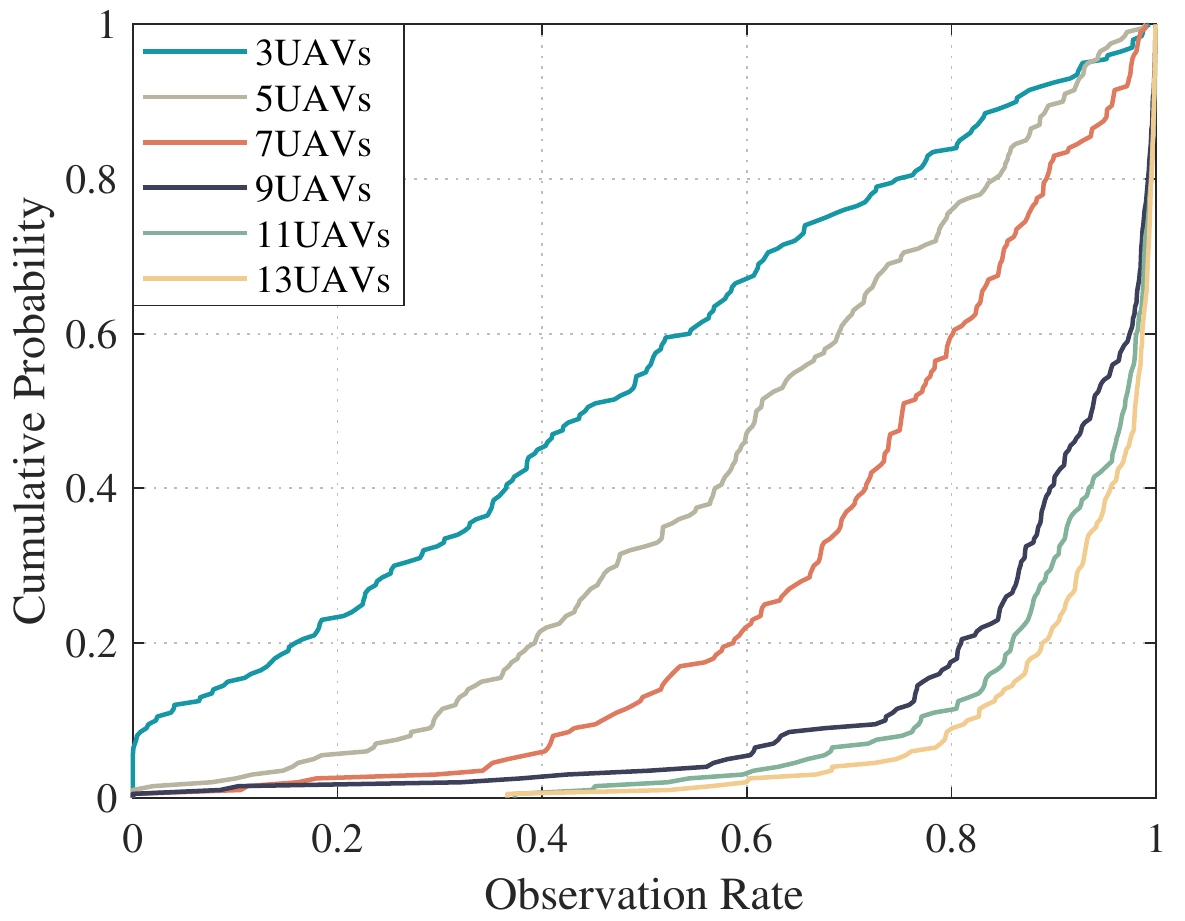} 
		\end{minipage}%
	}%

	\caption{The distribution of the observation coverage rate for different UAV numbers.}\label{fig13}
\end{figure}

\begin{figure}[htbp]
	\centering
	\begin{minipage}[t]{0.48\textwidth}
		\centering
		\includegraphics[width=1.85in,height=1.5in]{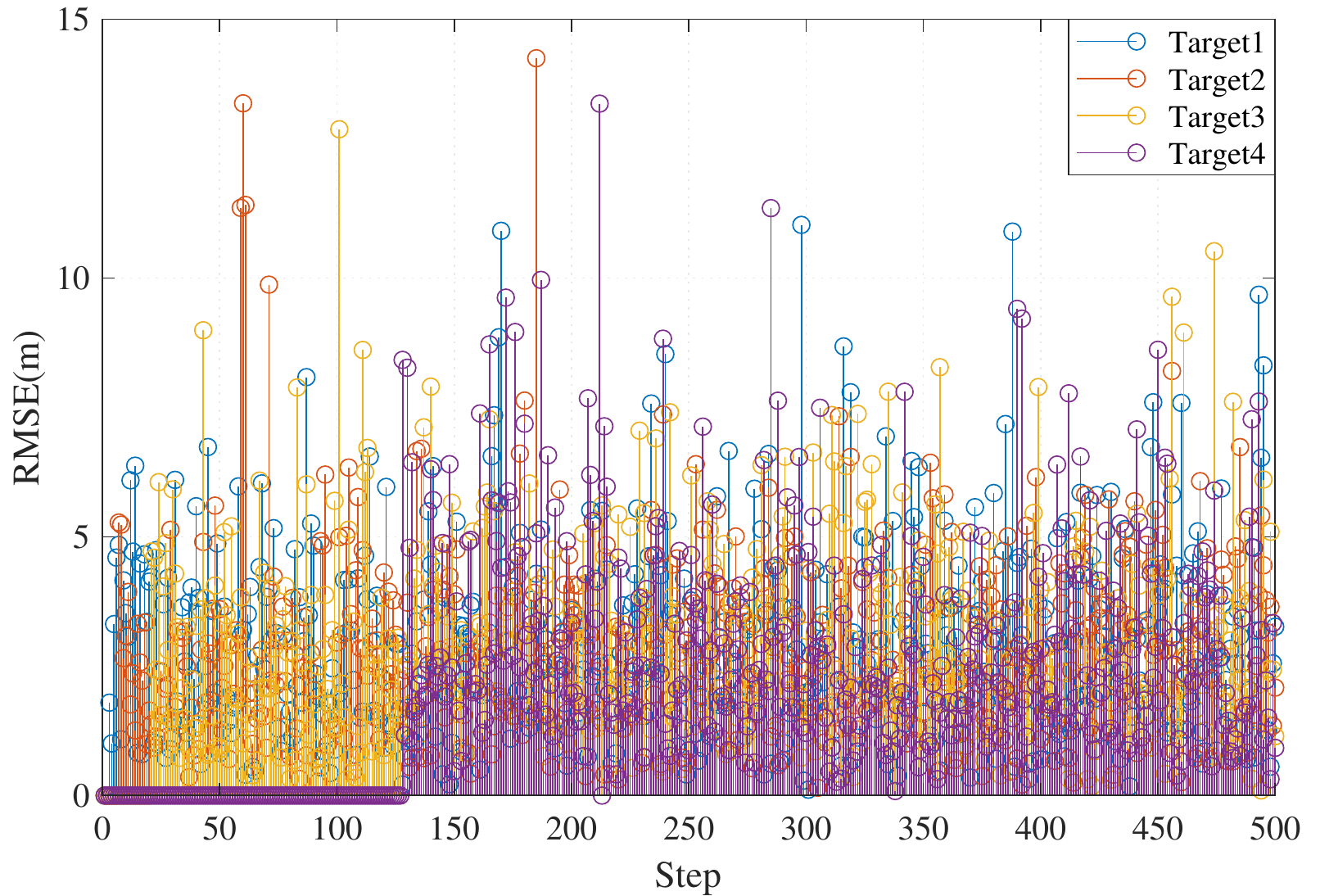} 
		\caption{The RMSE of targets in one simulation ($0$ indicates that the target is not detected or lost yet).}\label{fig14}
	\end{minipage}
	\begin{minipage}[t]{0.48\textwidth}
		\centering
		\includegraphics[width=1.85in,height=1.5in]{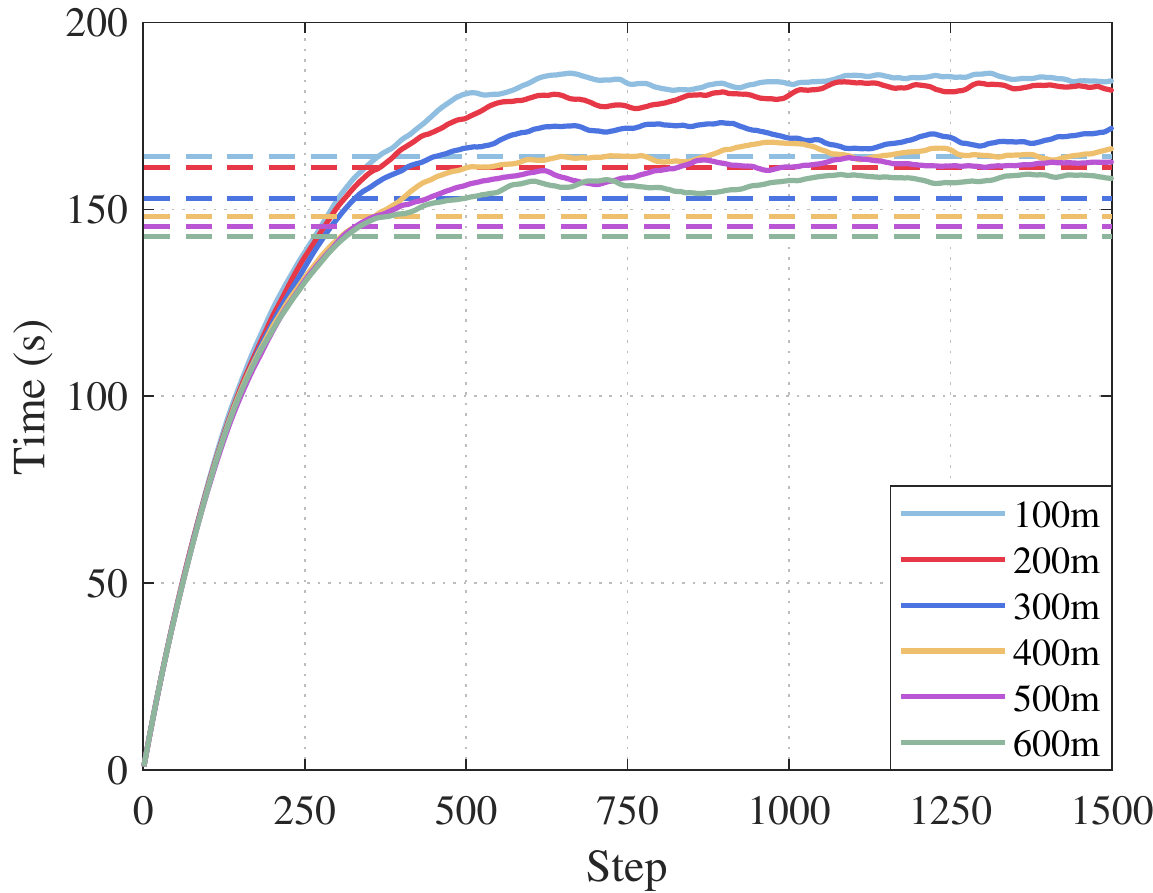} 
		\caption{The average uncovered time of different communication ranges with seven UAVs performing only coverage task.}\label{fig15}
	\end{minipage}
\end{figure}

For the target tracking effect, we use the observation coverage rate to represent the ratio of the time duration that the target is observed to the total simulation time. The average observation coverage rate of all targets is shown in Fig.~\ref{fig12}. Since the initial positions of the UAVs and the targets are random and the target moves randomly, the tracking performance can be measured directly using the overall average observation rate. 
Fig.~\ref{fig12} shows that with the increase of the UAV numbers, the average observation rate of the targets is also gradually increasing.
The distribution of the observation coverage rate for our repeated simulations is shown in Fig.~\ref{fig13}. It indicates that the observation coverage rate of most of the targets is maintained at a high level when the number of UAVs is more than that of targets. Moreover, Fig.~\ref{fig14} shows the RMSE of targets in one simulation.


\begin{figure}[htbp]%
	\centering
	\subfigure[]{
		\begin{minipage}[t]{0.48\linewidth}
			\centering
			\includegraphics[width=1.85in,height=1.5in]{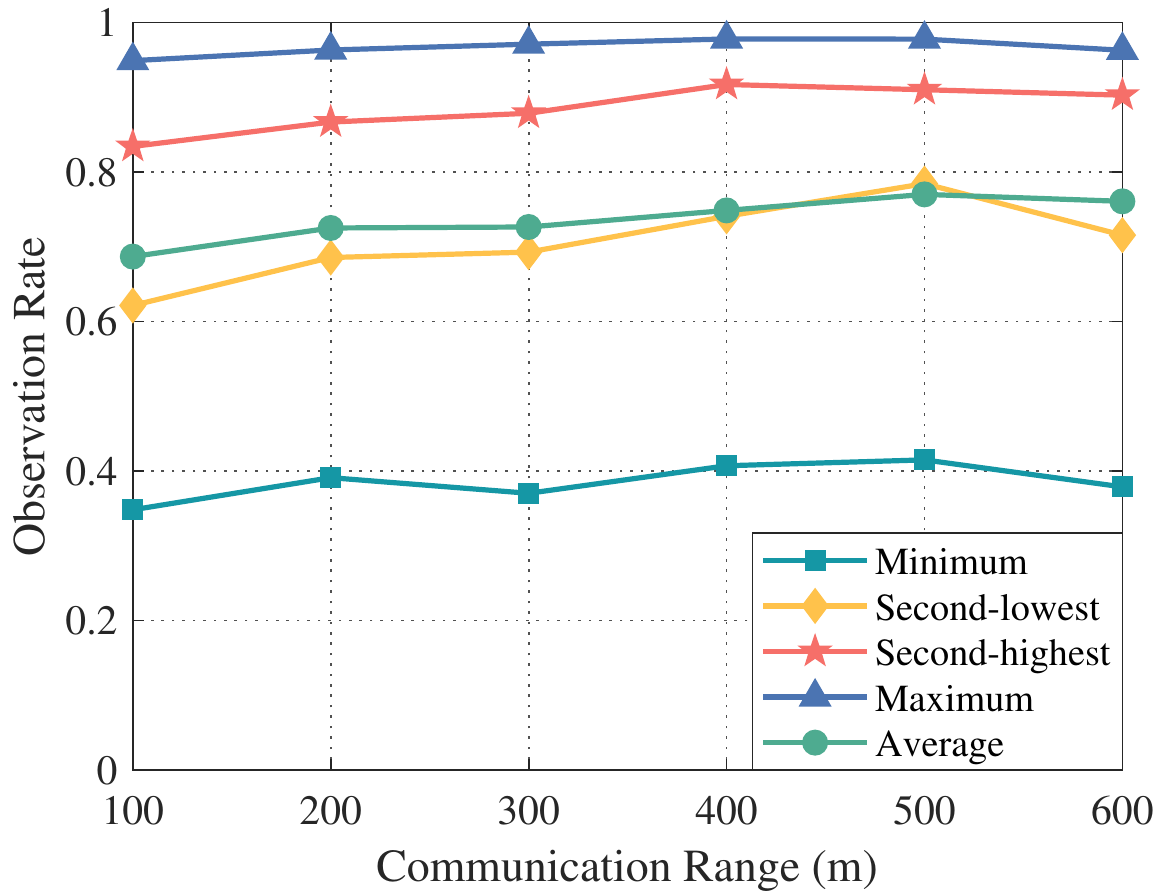}
		\end{minipage}%
	}%
	\subfigure[]{
		\begin{minipage}[t]{0.48\linewidth}
			\centering
			\includegraphics[width=1.85in,height=1.5in]{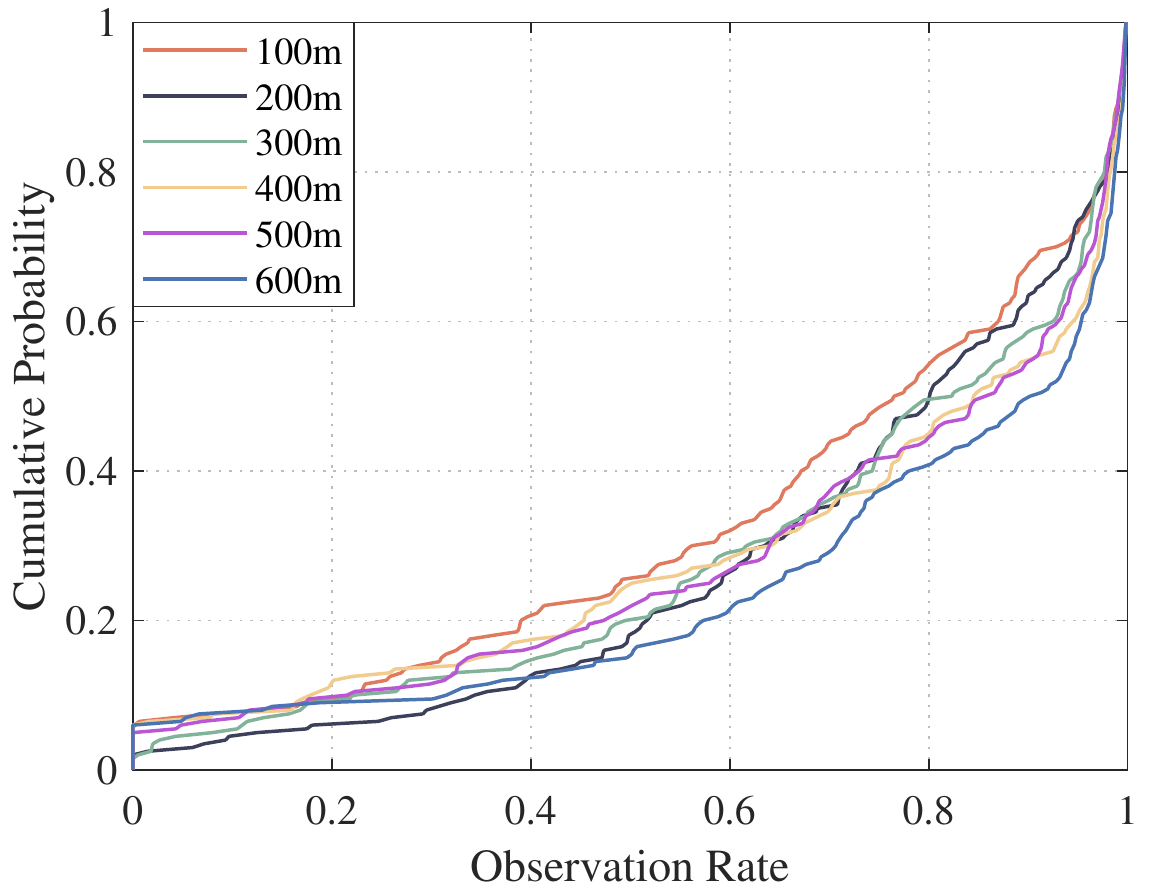}
		\end{minipage}%
	}%
	
	\caption{The observation coverage rate of different communication ranges with nine UAVs and four targets. (a) is the observation coverage rate from lowest to highest and the average observation coverage rate of targets, (b) is the distribution of the observation coverage rate.}\label{fig16_1}
\end{figure}

Then, we evaluate the performance under different communication ranges, which vary from $100m$ to $600m$ (See Fig.~\ref{fig15} and~\ref{fig16_1}). To explore the influence of the communication range on the coverage performance, we consider that UAVs only perform the coverage task.
Fig.~\ref{fig15} indicates that the uncovered time decreases gradually with the increased communication ranges.
When simultaneously considering the search and tracking tasks, since the target tracking algorithm can ensure the targets are tracked most time,  the communication range has little effect on the observation coverage rate, as shown in Fig.~\ref{fig16_1}. 
Simulation results show that the area coverage and target tracking performance can be enhanced by improving communication capability.

	

\subsubsection{Comparison with other algorithms}\label{sec7_22}
Here, we compare our method with the simultaneous coverage and tracking (SCAT) algorithm in \cite{bib2} and the multi-target tracking (MT) algorithm in \cite{bib34}. 
The SCAT method in \cite{bib2} translates the area coverage and target tracking problem to the problem of covering environments with time-varying density functions. However, the target tracking task is not explicitly considered in SCAT.
The MT method in \cite{bib34} uses the distributed partially observable Markov decision algorithm to make tracking decisions. It assumes that the targets are initially located in the field of view of the UAVs. 
In order to make the comparison reasonable, we combine our coverage algorithm with the MT algorithm for the subsequent comparison, called CMT.
In addition, the task assignment of the MT algorithm adopts multiple rounds of 1-to-1 assignment to solve the many-to-many assignment problem. There are no restrictions on the number of UAVs tracking the same target, resulting in UAVs selecting to perform the tracking task. We limit the number of 1-to-1 assignments in the task allocation part of CMT and call this algorithm as L-CMT.

We compare these algorithms from the target observation coverage rate and the area uncovered time. We set the number of UAVs as  7  and the number of targets as 4 in subsequent comparisons.

\begin{figure}[htbp]
	\centering
	\begin{minipage}[t]{0.48\textwidth}
		\centering
		\includegraphics[width=1.85in,height=1.5in]{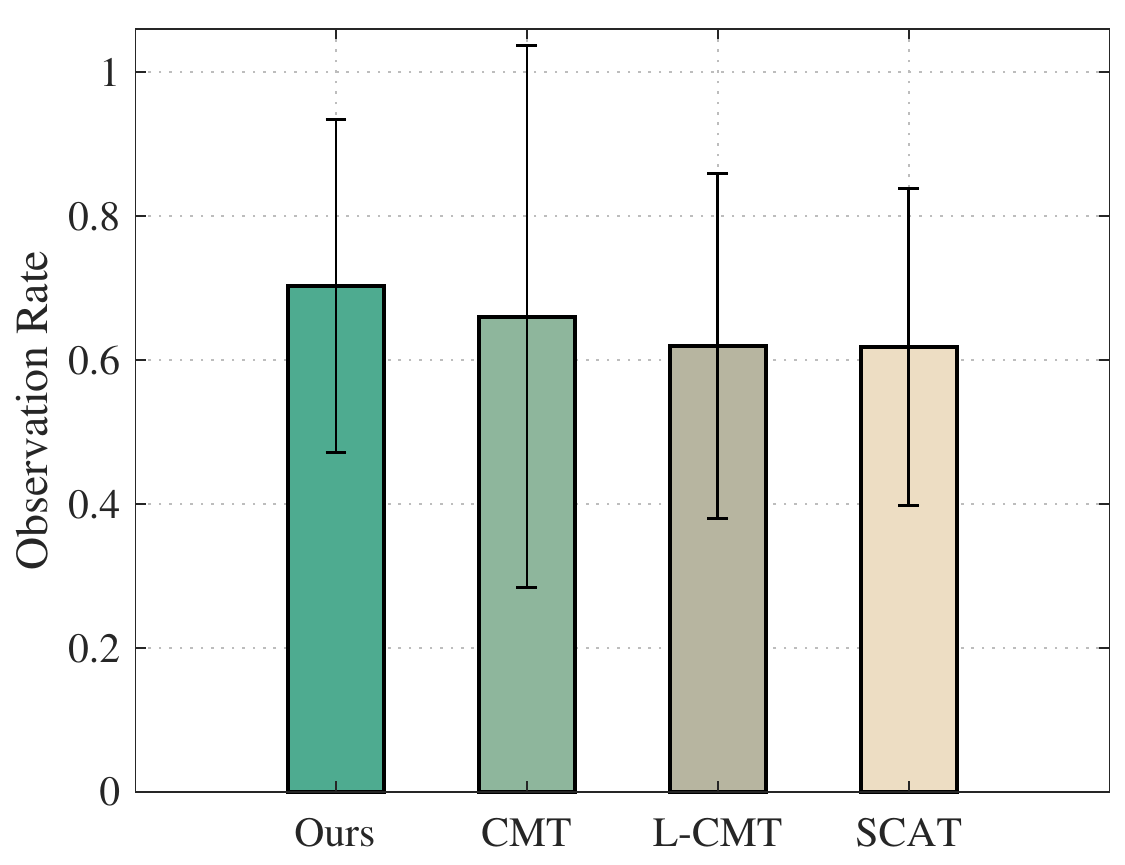}
		\caption{The average observation coverage rate of 7 UAVs and 4 targets for different methods.}\label{fig18}
	\end{minipage}
	\begin{minipage}[t]{0.48\textwidth}
		\centering
		\includegraphics[width=1.85in,height=1.5in]{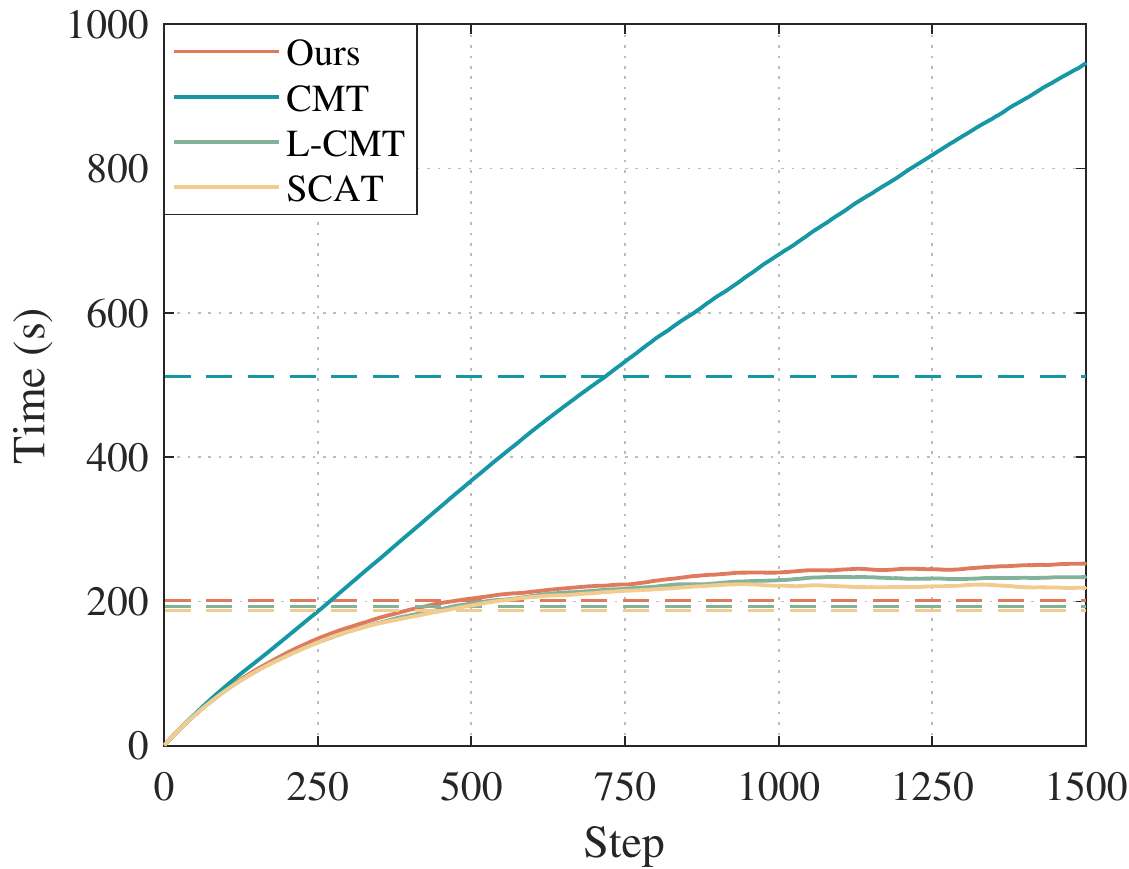}
		\caption{The average uncovered time of 7 UAVs and 4 targets for different methods.}\label{fig19}
	\end{minipage}
\end{figure}

Fig.~\ref{fig18} shows the average observation coverage rate of all targets. The average observation coverage rate is $0.7080 \pm 0.2311$ for our method, $0.6600 \pm 0.3761$ for the CMT method, $0.6190 \pm 0.2397$ for the L-CMT method, and $0.6177 \pm 0.2203$  for the SCAT method.
Fig.~\ref{fig19} shows the average uncovered time of the mission area. 
The average uncovered time of the SCAT method is the smallest, indicating the coverage performance of SCAT is the best.
Since the SCAT algorithm ignores the explicit tracking of the target, its coverage performance is good, while the tracking performance is poor.
The CMT algorithm does not limit the number of UAVs tracking the same target, making UAVs tend to choose the tracking task and sometimes focus on tracking the same targets. Thus, 
the CMT algorithm has the worst area coverage performance. Besides, the target observation coverage variance of CMT is the largest.
The L-CMT method improves the coverage performance of CMT by limiting the number of allocations. 
Due to the design of the task assignment minimum-cost maximum-flow algorithm, our method has the best tracking effect, and the area coverage effect is slightly lower than that of L-CMT and SCAT methods.
These simulation results indicate that our method is more suitable for the area coverage and target-tracking task.
\subsubsection{Other performance}\label{sec7_23}
Here, we also list the time taken by the allocation algorithm in Table~\ref{tab_3}, which shows that our allocation algorithm can be applied to the fast real-time allocation of large-scale tasks.
\begin{table}[h!]
	\begin{center}
		\begin{minipage}{0.85\textwidth}
			\caption{Running time of our allocation method} 
			\footnotesize
			\renewcommand{\arraystretch}{0.5}
			\setlength{\tabcolsep}{8mm}
			\begin{tabular}{@{}cccc@{}} 
				\toprule 
				\textbf{$ \bf N_u \leq \bf N_\tau$} & \textbf{$\bf 20V30$} & \textbf{$\bf 50V60$} & \textbf{$\bf 80V100$} \\ %
				\midrule 
				\textbf{Time use (s)} & 0.00218 & 0.00948 & 0.02526 \\   
				\midrule 
				\textbf{$\bf N_\tau < \bf N_u <  \bf \sum \nolimits_{j=1}^{N_\tau} {n_j}$} & \textbf{$\bf 60V50$} & \textbf{$\bf 80V50$} & \textbf{$\bf 100V50$} \\ %
				\midrule 
				\textbf{Time use (s)} & 0.00925 & 0.01168 & 0.01469 \\   
				\midrule 
 				\textbf{$\bf  N_u > \bf \sum \nolimits_{j=1}^{N_\tau} {n_j}$} & \textbf{$\bf 80V20$} & \textbf{$\bf 100V25$} & \textbf{$\bf 120V30$} \\ %
 				\midrule 
 				\textbf{Time use (s)} & 0.00491 & 0.00768 & 0.01083 \\  
 				\botrule
			\end{tabular}
			\label{tab_3} 
		\end{minipage}
	\end{center}
\end{table}

\begin{figure}[htbp]
	\centering
	\begin{minipage}[t]{0.48\textwidth}
	\centering
	\includegraphics[width=1.9in]{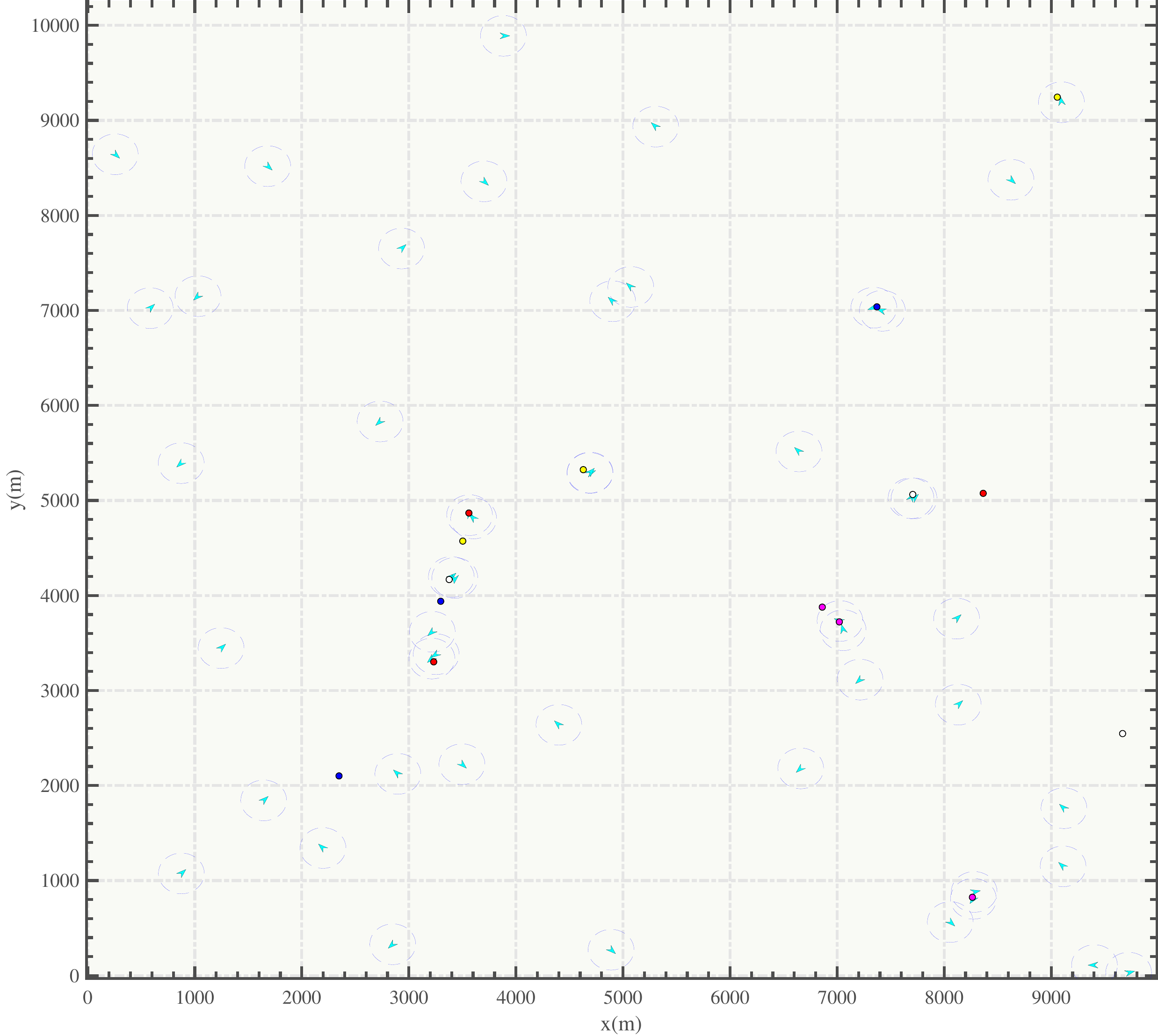}
	\caption{The area coverage and target tracking task with 50 UAVs and 15 targets in the area of 10km*10km.}\label{fig20}
	\end{minipage}
	\begin{minipage}[t]{0.48\textwidth}
	\centering
	\includegraphics[width=2.1in]{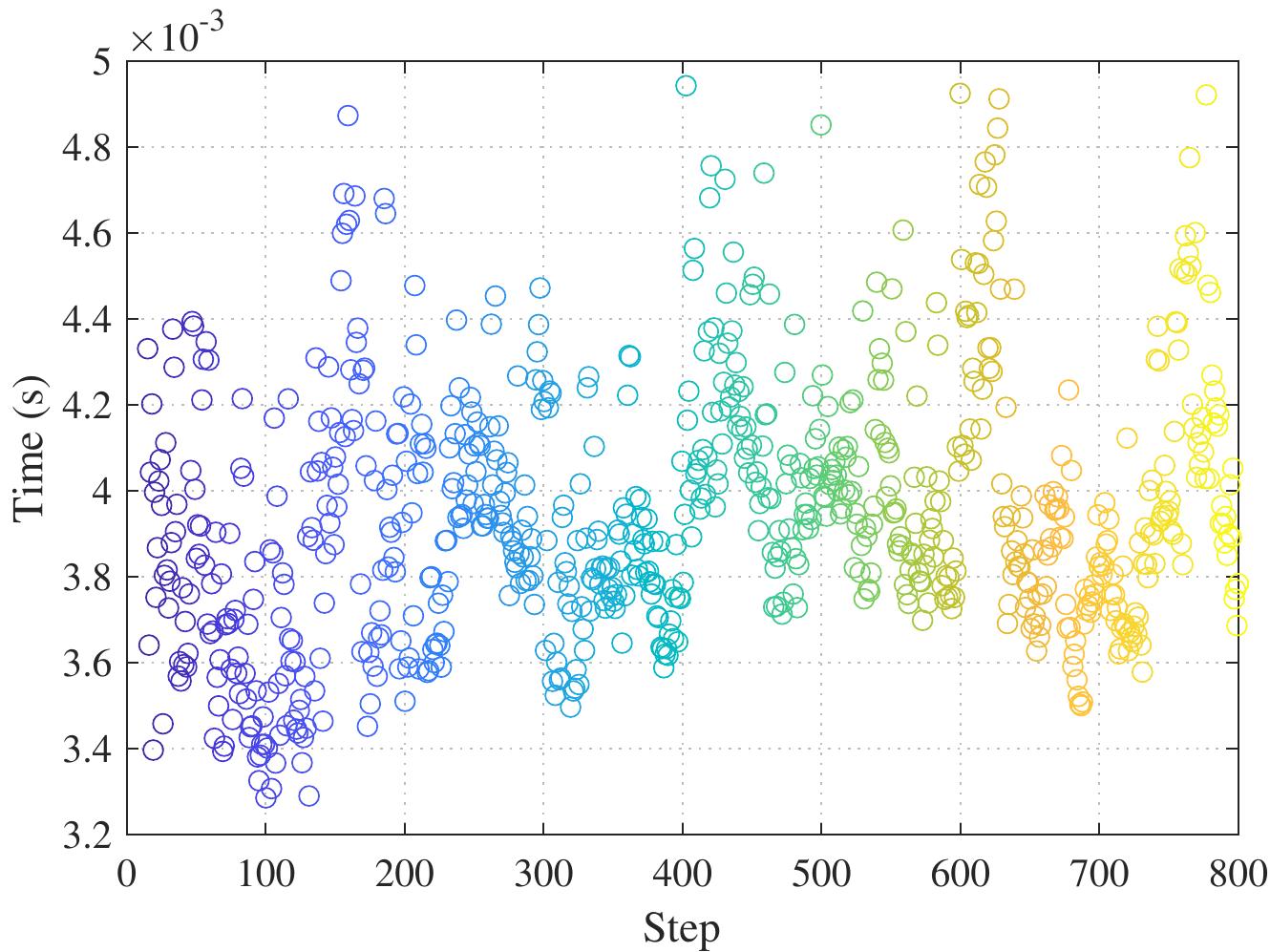}
	\caption{The single-step decision-making time in one simulation with 50 UAVs and 15 targets.}\label{fig21}
	\end{minipage}
\end{figure}

In addition, to reflect the scalability of the distributed algorithm, we set the task area as 10km*10km with 50 UAVs and 15 targets in Fig.~\ref{fig20}. Fig.~\ref{fig21} shows the single-step decision-making time of the simulation, which can meet the needs of real-time decision-making.

\subsection{Hardware-in-the-loop simulation}\label{sec7_3}
\begin{figure}[htbp]%
	\centering
	\subfigure[]{
		\begin{minipage}[t]{0.47\linewidth}
			\centering
			\includegraphics[width=1.6in]{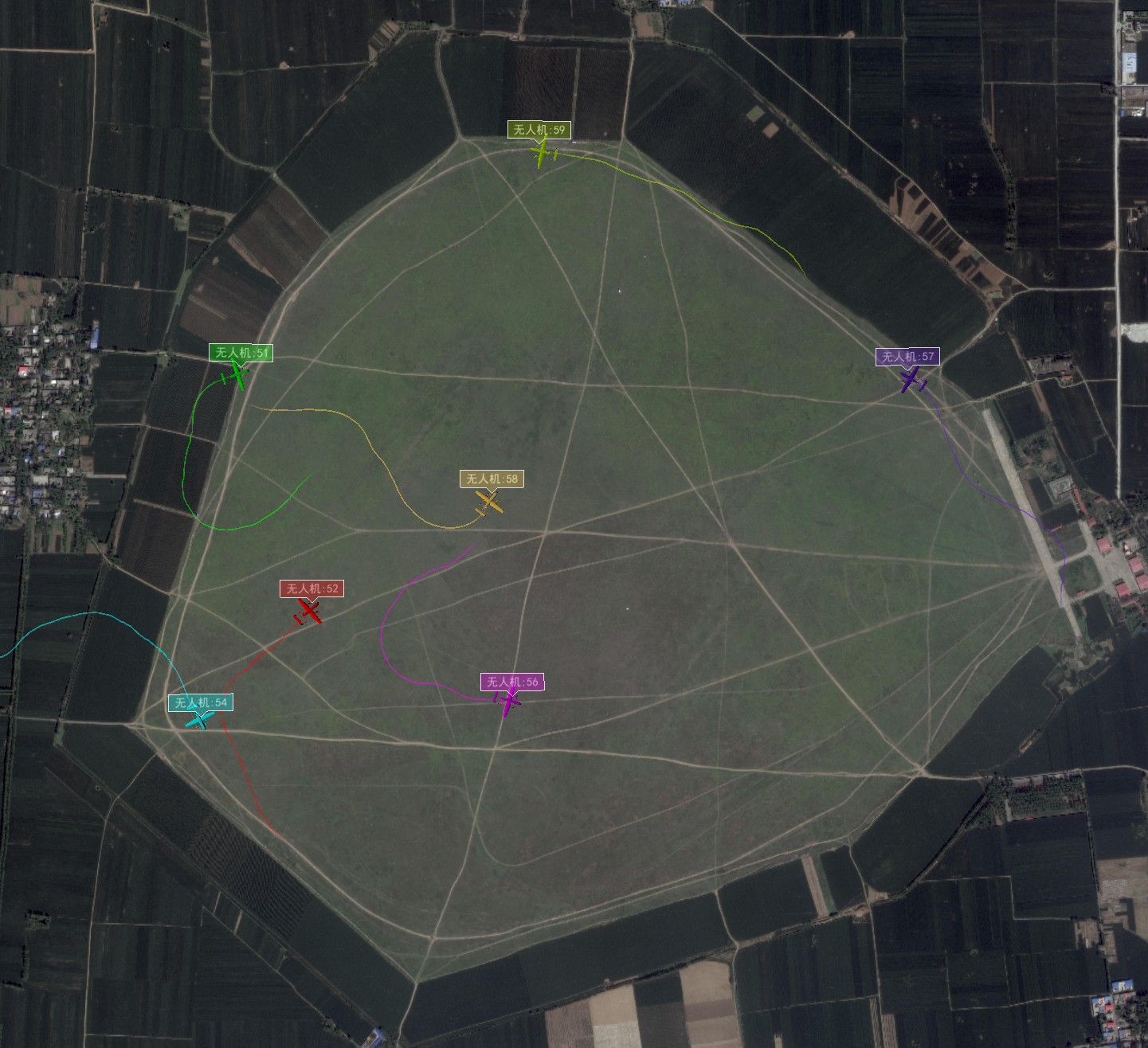}
			\label{fig22a}
		\end{minipage}%
	}%
	\subfigure[]{
		\begin{minipage}[t]{0.47\linewidth}
			\centering
			\includegraphics[width=1.6in]{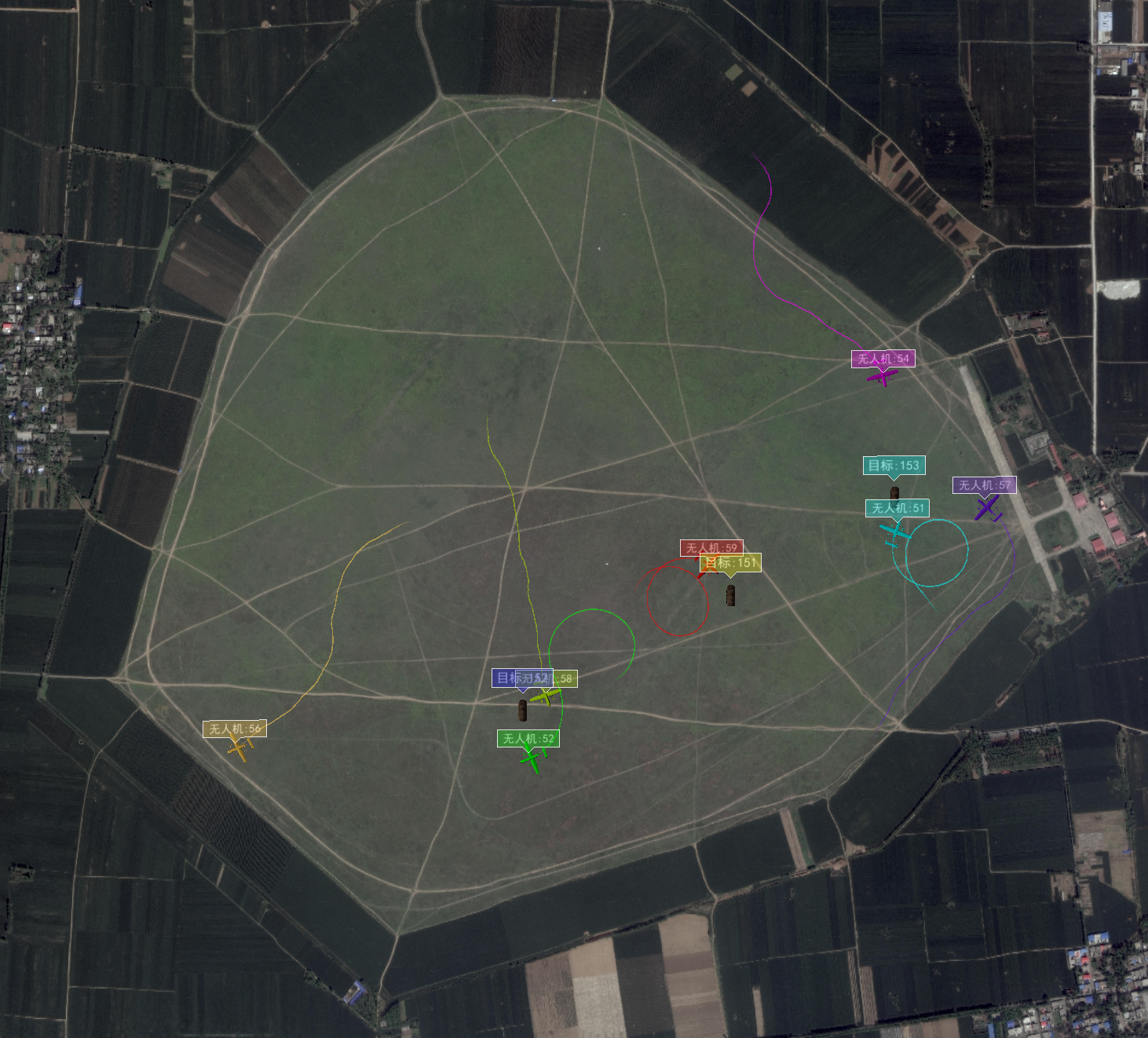}
			\label{fig22b}
		\end{minipage}%
	}%
	\caption{Hardware-in-the-loop simulation of the area coverage ans tracking task with 7 UAVs. (a) No targets, (b) 4 targets.}\label{fig22}
\end{figure}

The hardware-in-the-loop (HIL) verification of our distributed area coverage and target tracking algorithm is carried out through the the HIL simulator  \cite{bib55} constructed by our team.
We set the task area as 2.5km*2.5km with 7 UAVs and 4 moving targets. 
As shown in Fig.~\ref{fig22a}, all UAVs take the area coverage task when there are no targets. Fig.~\ref{fig22b} shows that the UAV automatically decides the task mode when detecting the targets, and each target is assigned to one UAV for tracking. While the remaining UAVs carry out the area coverage tasks.

\begin{figure}[htbp]
	\centering
	\begin{minipage}[t]{0.48\textwidth}
		\centering
		\includegraphics[width=1.85in,height=1.5in]{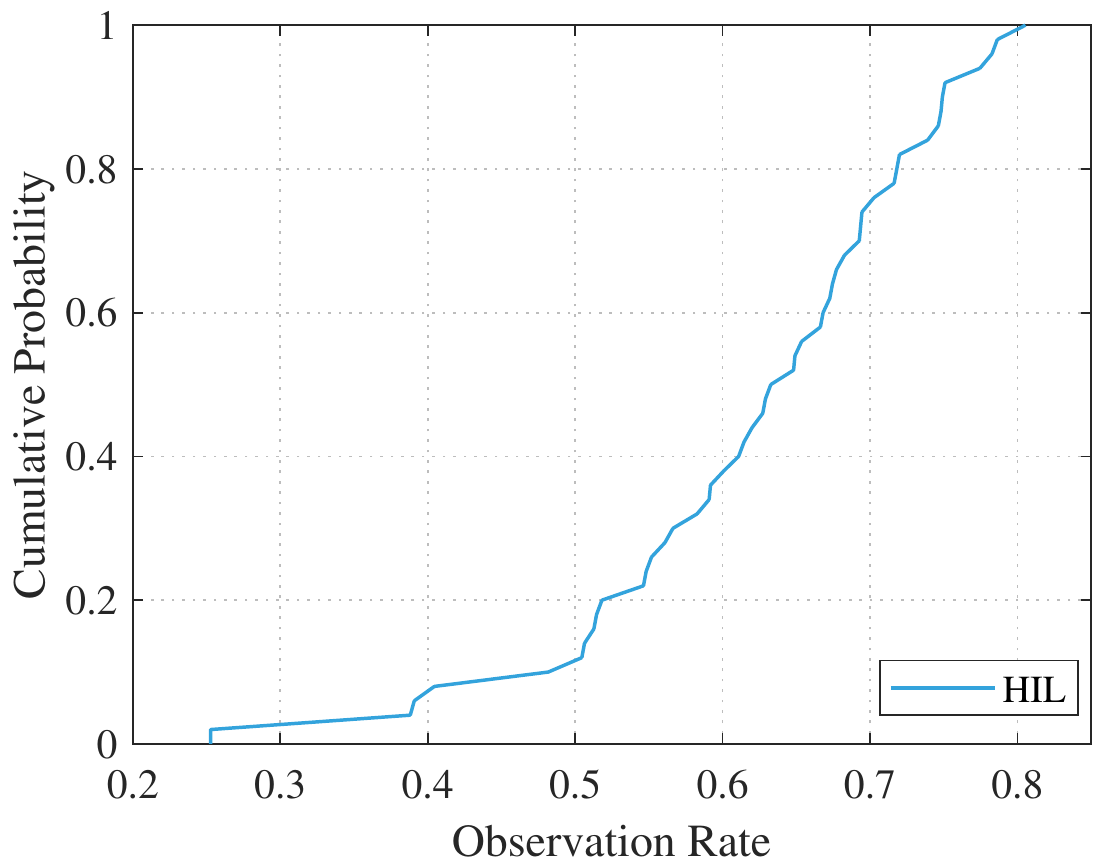}
		\caption{The average observation coverage rate of 20 HIL simulations.}\label{fig23}
	\end{minipage}
	\begin{minipage}[t]{0.48\textwidth}
		\centering
		\includegraphics[width=1.85in,height=1.5in]{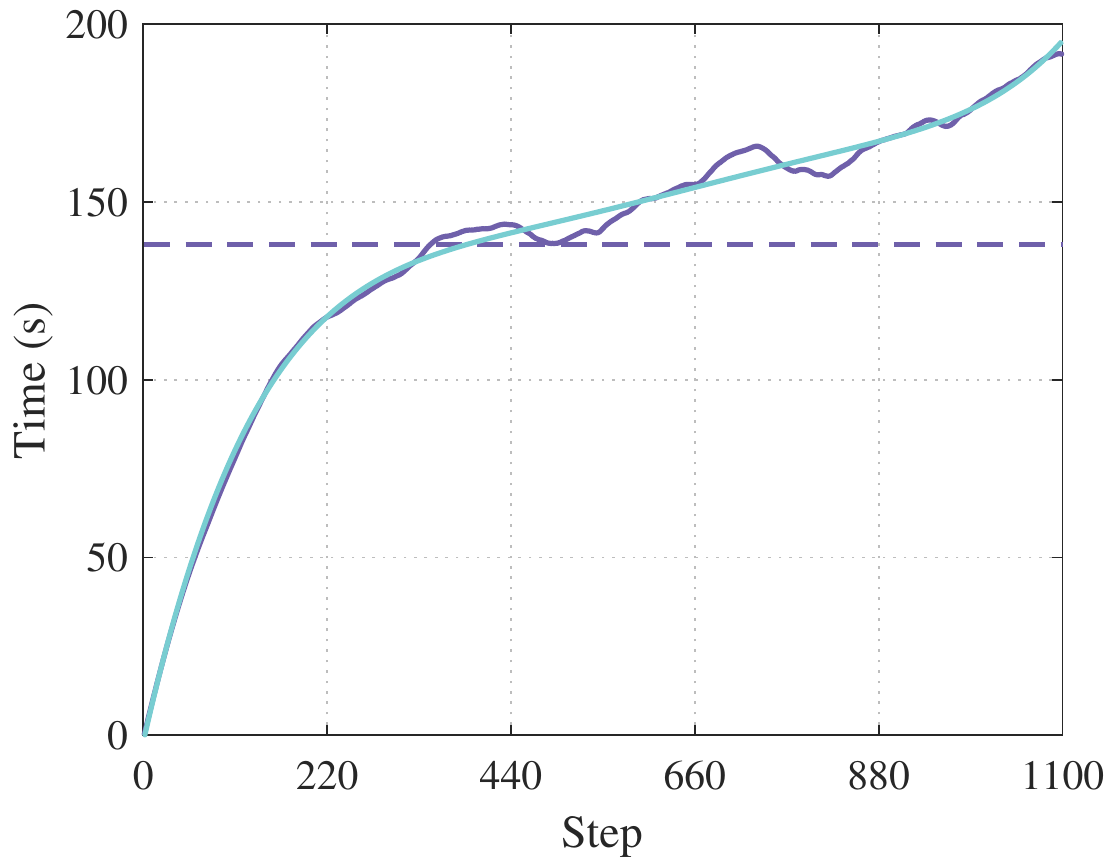}
		\caption{The average uncovered time of 20 HIL simulations.}\label{fig24}
	\end{minipage}
\end{figure}

\begin{figure}[htbp]%
	\centering
	\includegraphics[width=0.8\textwidth]{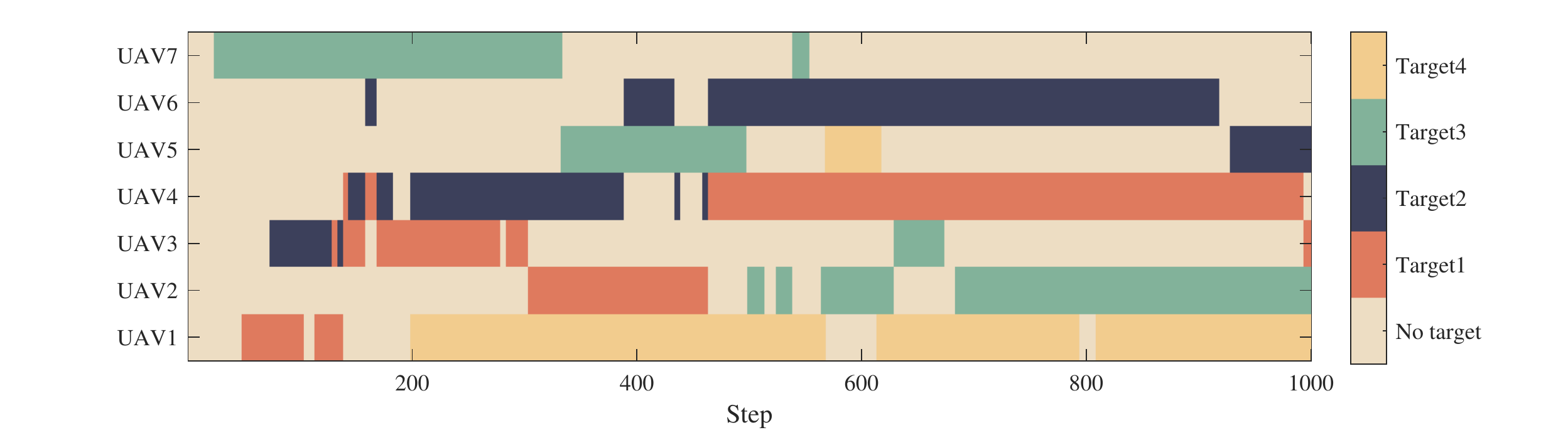}
	\caption{The allocation results in one HIL simulation.}\label{fig25}
\end{figure}

We count the target observation coverage rate and the area uncovered time of multiple HIL simulations (See Fig.~\ref{fig23} and~\ref{fig24}). 
Here, the coverage and tracking performance difference between HIL simulations and numerical simulations mainly comes from the uncertainty of the underlying modeling in hardware-in-the-loop simulations. The average observation coverage rate is $0.6240 \pm 0.1153$ in HIL simulatons. 
Moreover, the assignment results in one of the HIL simulations are given in Fig.~\ref{fig25}. The hardware-in-the-loop simulations demonstrate the feasibility of the real-time onboard implementations of our area coverage and target tracking algorithm.
\section{Conclusion}\label{sec8}
This paper proposes a distributed algorithm for area coverage and multi-target tracking. We use a distributed information fusion strategy based on maximum consensus protocol to estimate the joint state of multi-targets and get the joint detection information of the mission area. We also design a task allocation algorithm based on minimum cost and maximum flow. Optimal tracking and area coverage action are obtained based on optimal planning and distributed anti-flocking algorithm. In the distributed information fusion phase, we introduce the area compression map to extend the coverage algorithm to any large task area, and the scalability is verified in the simulation. In addition, combined with the consensus strategy, the designed network flow allocation algorithm is implemented in a distributed way. And we use the Fisher information as the task reward of target tracking.

We apply the integrated algorithm to area coverage and target tracking tasks and study the effects of the number of UAVs and communication range on the coverage and tracking performance. The results show that the coverage performance of UAVs can be significantly enhanced by increasing their scale and improving their communication capability. However, the communication range has little effect on the tracking performance. In addition, we apply our algorithm in the hardware-in-the-loop simulation of area coverage and target tracking tasks.

 In future work, we will consider these problems and focus on reducing the communication load and dealing with the communication delay problem to promote the practical application of the algorithm.








\bibliography{refe.bib}

\end{document}